\documentclass[fleqn,usenatbib]{mnras}

\usepackage{newtxtext,newtxmath}
\usepackage[T1]{fontenc}
\usepackage{ae,aecompl}

\usepackage{graphicx}	% Including figure files
\usepackage{amsmath}	% Advanced maths commands
\usepackage{amssymb}	% Extra maths symbols
\usepackage{siunitx}    % SI unit formatting
\usepackage{multirow}   % spanning multiple rows/columns in tables
\usepackage{breakcites} % aids in breaking long citations over lines without error
 
\newcommand{\SIadj}[2]{\SI[number-unit-product={\text{-}}]{#1}{#2}}

\DeclareSIUnit{\solarmass}{M_\odot}
\newcommand{\pp}{{\sc ppmap}}
\newcommand{\ppp}{{\sc ppmap}$\;$}
\newcommand{\pacs}{{\sc pacs}}
\newcommand{\pacss}{{\sc pacs}$\;$}

\newcommand{\spiree}{{\sc spire}$\;$}
\newcommand{\rwl}{_{_{300}}}
\newcommand{\subB}{_{_{\rm B}}}
\newcommand{\subC}{_{_{\rm C}}}
\newcommand{\subD}{_{_{\rm D}}}

\newcommand{\subO}{_{_{\rm O}}}
\newcommand{\subP}{_{_{\rm P}}}
\newcommand{\subS}{_{_{\rm S}}}

\newcommand{\unc}{\varSigma}
\title[A \pp~view of L1495]{L1495 Revisited: A \pp~View of a Star-Forming Filament}
\author[A. D. P. Howard et al.]{
A. D. P. Howard,$^{1}$\thanks{E-mail: Alexander.Howard@astro.cf.ac.uk (CU)}
A. P. Whitworth,$^{1}$
K. A. Marsh,$^{3}$
S. D. Clarke,$^{2}$
M. J.Griffin$^{1}$
\newauthor
M. W. L. Smith$^{1}$ and O. D. Lomax$^{4}$
\\
$^{1}$School of Physics and Astronomy, Cardiff University, 5 The Parade, Cardiff, CF24 3AA, UK\\
$^{2}$I. Physikalisches Institut, Universit{\"a}t zu K{\"o}ln, Z{\"u}lpicher Str. 77, D-50937 K{\"o}ln, Germany\\
$^{3}$IPAC, Caltech, 1200E California Boulevard, Pasadena, CA 91125, USA\\
$^{4}$ESTEC, Keplerlaan 1, 2201 AZ Noordwijk, Netherlands
}

\date{Accepted XXX. Received YYY; in original form ZZZ}

\pubyear{2019}

\begin{document}
\label{firstpage}
\pagerange{\pageref{firstpage}--\pageref{lastpage}}
\maketitle

\defcitealias{Palmeirim2013}{P13}

%%%%%  Abstract of the paper %%%%% [W=223]
\begin{abstract}
We have analysed the {\em Herschel} and {\sc SCUBA-2} dust continuum observations of the main filament in the Taurus L1495 star forming region, using the Bayesian fitting procedure \pp. (i) If we construct an average profile along the whole length of the filament, it has {\sc fwhm} $\simeq 0.087\pm 0.003\,{\rm pc};\;$, but the closeness to previous estimates is coincidental. (ii) If we analyse small local sections of the filament, the column-density profile approximates well to the form predicted for hydrostatic equilibrium of an isothermal cylinder. (iii) The ability of \ppp to distinguish dust emitting at different temperatures, and thereby to discriminate between the warm outer layers of the filament and the cold inner layers near the spine, leads to a significant reduction in the surface-density, $\varSigma$, and hence in the line-density, $\mu$. If we adopt the canonical value for the critical line-density at a gas-kinetic temperature of $10\,{\rm K}$, $\mu_{_{\rm CRIT}}\simeq 16\,{\rm M_{_\odot}\,pc^{-1}}$, the filament is on average trans-critical, with ${\bar\mu}\sim \mu_{_{\rm CRIT}};\;$ local sections where $\mu >\mu_{_{\rm CRIT}}$ tend to lie close to pre-stellar cores. (iv) The ability of \ppp to distinguish different types of dust, i.e. dust characterised by different values of the emissivity index, $\beta$, reveals that the dust in the filament has a lower emissivity index, $\beta\la1.5$, than the dust outside the filament, $\beta\ga 1.7$, implying that the physical conditions in the filament have effected a change in the properties of the dust.
\end{abstract}

% Select between one and six entries from the list of approved keywords. Don't make up new ones.

%%%%%%%%%%%%%%%%% BODY OF PAPER %%%%%%%%%%%%%%%%%%

%%%%%
\section{Introduction}\label{SEC:Int}
%%%%%

Filaments appear to be critical structures in the star formation process, linking the molecular cloud scale, $\ga 3\,{\rm pc}$, to the core scale, $\la 0.3\,{\rm pc}$, \citep{Andre2010,Arzoumanian2011,Hacar2013,Konyves2015,Marsh2016}. Consequently there have been many studies aimed at understanding the formation of filaments, and their evolution and fragmentation into cores, both from an observational perspective \citep[e.g.][]{Andre2010, Arzoumanian2011, Hacar2013, Kainulainen2013, Panopoulou2014, Konyves2015, Tafalla2015, Andre2016, Cox2016, Kainulainen2016, Marsh2016, Kainulainen2017, Hacar2018} and from a theoretical perspective \citep[e.g.][]{Ostriker1964, Inutsuka1992, Inutsuka1997, Fischera2012, Heitsch2013, Smith2014, Clarke2015, Seifried2015, Clarke2016, Gritschneder2016, Smith2016, Clarke2017, Seifried2017, Clarke2018, Heigl2018}.

The L1495 filament in Taurus has been extensively studied as a site of low- and intermediate-mass star formation \citep{Shu1987, Strom1994, Nakamura2008, Hacar2013, Palmeirim2013, Seo2015, Tafalla2015, Marsh2016, WardThompson2016, Punanova2018}, due to its proximity \citep[distance, $D\,$\SI{\sim 140}{pc};][]{Elias1978}, its large physical size on the sky (lateral extent, $L\sim 4\,{\rm pc}$), and the fact that there is little evidence for vigorous feedback from nearby high-mass stars. 

Using {\em Herschel}$\,$\footnote{{\em Herschel} is  an  ESA  space  observatory  with  science instruments provided by European-led Principal Investigator consortia and with important participation from NASA.} observations of thermal dust emission, \citet[][hereafter \citetalias{Palmeirim2013}]{Palmeirim2013} estimate that the width of the L1495 filament is $\sim 0.1\,{\rm pc}$. This is the characteristic filament width seen in many local star forming regions by \citet{Arzoumanian2011}, although \citet{Panopoulou2017} have argued that this is an artefact of the procedure used to determine filament widths. \citetalias{Palmeirim2013} also find that the L1495 filament is thermally super-critical throughout most of its length, i.e. its line-density is too large for it to be supported against radial collapse by a thermal pressure gradient. There are several pre-stellar cores \citep[e.g.][]{Onishi2002,Marsh2016} and protostellar objects \cite[e.g.][]{Motte2001,Rebull2010} embedded in the L1495 filament, suggesting that it has fragmented longitudinally --- although \citet{Schmalzl2010} point out that a large section of the filament, designated B211, contains no cores. 

In this paper we re-analyse {\em Herschel} and SCUBA-2 observations of the L1495 filament, using a new version of the Bayesian fitting algorithm \ppp \citep{Marsh2015}. Section \ref{SEC:Obs} describes the observations. Section \ref{SEC:approx} lists the approximations that are made and derives the factor for converting dust optical depths into column-densities of molecular hydrogen. Section \ref{SEC:Stand} reviews the standard procedure used previously to analyse maps of thermal dust emission in the far-infrared and submillimetre. Section \ref{SEC:PPMAP} outlines the new enhanced version of \pp, and the advantages it brings. Section \ref{SEC:prods} presents the raw data products obtained by applying \ppp to L1495. Section \ref{SEC:cyl} describes the methods used to analyse these data products in terms of a cylindrically symmetric model filament, and the results of this analysis. Section \ref{SEC:sub} addresses briefly the issue of internal sub-structure within the filament. Section \ref{SEC:disc} (a) compares synthetic maps generated using the results of our analysis with the original {\it Herschel} maps, and with maps generated using the results of previous analyses; and (b) shows that there is sufficient time for dust grains to accrete mantles in the interior of the filament. Section \ref{SEC:conc} summarises our main conclusions.

%%%%%
\section{Observations of L1495}\label{SEC:Obs}
%%%%%

%%%%%
\subsection{Herschel observations}
%%%%%

All but one of the maps used in our analysis are taken from four observations of the L1495 molecular cloud, performed as part of the {\em Herschel Gould Belt Survey} (HGBS)\footnote{\url{http://www.herschel.fr/cea/gouldbelt/en/}}. They comprise \SIadj{70}{\micro \metre} and \SIadj{160}{\micro \metre} data from \pacss \citep{Poglitsch2010}, plus  \SIadj{250}{\micro \metre}, \SIadj{350}{\micro \metre} and \SIadj{500}{\micro \meter} data from \spiree \citep{Griffin2010}, captured in the fast scan (\SI{60}{\arcsecond \per \second}) \pacs/\spiree parallel mode. The two nominal North-South scans were taken on 12 February 2010 and 7 August 2010, with the orthogonal East-West scan taken on 8 August 2010. A fourth scan, also in the nominal North-South direction, was taken on 20 March 2012, to target a small region not previously covered by the \pacss data. The {\em Herschel} Observation IDs for these scans are 1342202254, 1342190616, 1342202090 and 1342242047 respectively. 

The calibrated scans were reduced using the {\sc hipe} Continuous Integration Build (CIB) Number 16.0.194, which uses the finalised reduction pipelines. \pacss maps were produced using a modified version of the \verb|JScanam| task, whilst \spiree maps were produced with the \verb|mosaic| script operating on Level 2 data products. The \pacss observations were compared to {\em Planck} and {\sc iras} data to determine the sky offset values that should be applied \citep[cf.][]{Bernard2010}; the sky median values adopted for the \pacss \SIadj{70}{\micro \meter} and \SIadj{160}{\micro \meter} observations were 4.27 and 69.2 \si{MJy} respectively. Zero-point corrections for the \spiree observations were applied as part of the standard {\sc hipe} processing. 

The intrinsic angular resolutions, given as {\sc fwhm} beamsizes, are \SI{8.5}{\arcsecond}, \SI{13.5}{\arcsecond}, \SI{18.2}{\arcsecond}, \SI{24.9}{\arcsecond}, and \SI{36.3}{\arcsecond}, for, respectively, the \SIadj{70}{\micro \meter}, \SIadj{160}{\micro \meter}, \SIadj{250}{\micro \meter}, \SIadj{350}{\micro \meter}, and \SIadj{500}{\micro \meter} wavebands \citep{PACShbook,SPIREhbook}. The fast scan speed distorts the \pacss beams, giving effective beamsizes of $\sim 6\si{\arcsecond} \times 12\si{\arcsecond}$ for the \SI{70}{\micro \meter} waveband, and $\sim 12\si{\arcsecond} \times 16\si{\arcsecond}$ for the \SI{160}{\micro \meter} waveband; the values quoted above for \SI{70}{\micro \meter} and \SI{160}{\micro \meter} are angle-averaged means.

%%%%%
\subsection{SCUBA-2 observations}
%%%%%

We supplement the {\em Herschel~} observations with SCUBA-2 \citep{Holland2013} \SIadj{850}{\micro \meter} observations, taken as part of the JCMT Gould Belt Survey \citep{Ward-Thompson2007}. The intrinsic angular resolution of SCUBA-2 at \SI{850}{\micro \meter} is \SI{14.6}{\arcsecond}. The observations consist of \SI{30}{\arcminute} diameter circular regions made using the PONG1800 mapping mode \citep{Chapin2013}. Individual regions are mosaicked together. Full reduction of the SCUBA-2 observations is described in \citet{Buckle2015}.

As the SCUBA-2 processing of the L1495 field uses a high-pass filter set to \SI{10}{\arcminute} (in order to remove the effects  of atmospheric and instrumental noise), emission from large angular  scales is suppressed. To restore the larger spatial scales we combine  the SCUBA-2 map with an \SIadj{850}{\micro \meter} map from Planck, using the CASA {\sc  feather} task \citep{McMullin2007}. An optimised python  script to implement this combination is being written and will be released in Smith et al. (in prep.).

%%%%%
\section{Approximations}\label{SEC:approx}
%%%%%

%%%%%
\subsection{Opacity Law}
%%%%%

We follow the convention of parametrising the variation of the mass opacity coefficient (per unit mass of dust and gas), $\kappa_{_\lambda}$, with wavelength, $\lambda$, using an emissivity index,
\begin{eqnarray}
\beta&=&-\;\left.\frac{d\ln\left(\kappa_{_\lambda}\right)}{d\ln(\lambda)}\right|_{\lambda_{_{\rm o}}}\,,
\end{eqnarray}
where $\lambda\subO$ is an arbitrary reference wavelength. Here we use $\lambda\subO\!=\;$\SI{300}{\micro\meter}. It follows that, if the opacity at \SI{300}{\micro\meter} is $\kappa\rwl$,  the opacity at other nearby wavelengths can be approximated by
\begin{eqnarray}
\kappa_{_\lambda}&\simeq&\kappa\rwl\,\left(\!\frac{\lambda}{\SI{300}{\micro\meter}}\!\right)^{\!-\beta}\,.
\end{eqnarray}

%%%%%
\subsection{Conversion factors}\label{SEC:ConvFact}
%%%%%

The optical depth at \SI{300}{\micro\meter}, $\tau\rwl$, is related to the surface-density of dust, $\varSigma\subD$, by $\tau\rwl = \varSigma\subD\kappa\rwl$, so
\begin{eqnarray}
\varSigma\subD&=&\frac{\tau\rwl}{\kappa\rwl}\,.
\end{eqnarray}
The surface-density of dust, $\varSigma\subD$, is related to the total surface-density, $\varSigma$ (i.e. dust plus gas), by $\varSigma\subD=Z\subD\varSigma$, where $Z\subD$ is the fractional abundance of dust by mass, so
\begin{eqnarray}
\varSigma&=&\frac{\varSigma\subD}{Z\subD}\;\,=\;\,\frac{\tau\rwl}{Z\subD\,\kappa\rwl}\,.
\end{eqnarray}
If the hydrogen is totally molecular, and the fractional abundance of hydrogen by mass is $X$, then the column-density of molecular hydrogen is given by
\begin{eqnarray}
N_{_{{\rm H}_2}}&=&\frac{X\,\varSigma}{2\,m_{_{\rm H}}}\;\,=\;\,\frac{X\,\tau\rwl}{2\,m_{_{\rm H}}\,Z\subD\,\kappa\rwl}.
\end{eqnarray}

We stress that the fundamental quantity obtained from the analysis of {\em Herschel} maps is the dust optical-depth. However, it is easier to evaluate the results in terms of the associated total surface-density, $\varSigma$, or the associated column-density of molecular hydrogen, $N_{_{{\rm H}_2}}\!.$ For this purpose, we take the fractional abundance by mass of hydrogen to be $X=0.70$, the fractional abundance by mass of dust to be $Z\subD =0.01$, and the dust absorption opacity at \SI{300}{\micro\meter} to be $\kappa\rwl = 10\,\rm{cm^2\,g^{-1}}$. With these values -- and presuming the gas and dust are co-extensive -- we have
\begin{eqnarray}\label{EQN:tau2NH2}
N_{_{{\rm H}_2}}&=&\left[2.1\times 10^{24}\;{\rm H}_{_2}\,{\rm cm}^{-2}\right]\,\tau\rwl\,,\\\label{EQN:tau2Sigma}
\varSigma&=&\left[4.8\times 10^4\,{\rm M}_{_\odot}\,{\rm pc}^{-2}\right]\,\tau\rwl\,,\\
&=&N_{_{\rm H_2}}{\bar m}_{_{\rm H_2}}\,,\\\label{EQN:mbarH2}
{\bar m}_{_{\rm H_2}}&=&2m_{_{\rm H}}/X\;\,=\;\,\left[4.77\times 10^{-24}\,{\rm g\,H_{_2}^{-1}}\right]\,;
\end{eqnarray}
${\bar m}_{_{\rm H_2}}$ is the mass associated with one hydrogen molecule, when account is taken of other species, in particular helium. Eqns. (\ref{EQN:tau2NH2}) through (\ref{EQN:mbarH2}) will be used throughout the paper to convert dust optical depths into total surface-densities, $\varSigma$, and column-densities of molecular hydrogen, $N_{_{\rm H_2}}$.

%%%%%
\subsection{Caveats}
%%%%%

The factors in square brackets in Eqns. (\ref{EQN:tau2NH2}) and (\ref{EQN:tau2Sigma}) are {\it not} accurate to two significant figures. Moreover, when we derive variations in the emissivity index, $\beta$ (see Sections \ref{SEC:PPMAP} through \ref{SEC:cyl}), we should be mindful that these variations are almost certainly due to grain growth and/or coagulation, and therefore are likely to be accompanied by correlated changes in (i) the abundance by mass of dust, $Z\subD$, and (ii) the dust absorption opacity at the reference wavelength, $\kappa\rwl$. The magnitudes of these changes are not currently known, and even their sense is not established with total certainty. This uncertainty does not affect the variations in $\beta$ which we detect, only the amount of mass ($\varSigma$) or molecular hydrogen ($N_{_{{\rm H}_2}}$) associated with the different types of dust. Thus, when we refer to the line-of-sight mean emissivity index, ${\bar\beta}$ (e.g. Eqn. \ref{EQN:beta_bar}), or the line-of-sight mean temperature, $\bar{T}\subD$ (e.g. Eqn. \ref{EQN:T_bar}), we should be mindful that these are strictly speaking optical depth weighted means, and only approximately mass-weighted means. In contrast, the means returned by the standard procedure (see Section \ref{SEC:Stand}) are flux weighted means.

%%%%%
\section{The standard procedure for analysing maps of thermal dust emission}\label{SEC:Stand}
%%%%%

The standard procedure for analysing far-infrared and submillimetre maps of thermal dust emission proceeds by smoothing all maps to the coarsest resolution, which for the {\em Herschel} maps used here means the resolution at the longest wavelength (\SI{500}{\micro\meter}), i.e.  $\sim 36.3$\si{\arcsecond}. A large amount of information is lost when the shorter-wavelength maps are smoothed. In certain cases, spatial filtering techniques have been applied to increase the final resolution. For example, \citetalias{Palmeirim2013} are able to recover the \spiree \SI{250}{\micro\meter} resolution, $\sim\!18$\si{\arcsecond}. However, with {\em Herschel} observations this is the limit \citep{Andre2010}, so the finer resolution of the  \pacss wavebands is still lost.

Next, the standard procedure assumes that the dust on the line of sight through each pixel is of a single type and at a single temperature. In other words, the dust emissivity index, $\beta$, and the dust temperature, $T\subD$, are taken to be uniform along the line of sight. This is a very crude assumption. There is growing evidence that the properties of dust evolve towards different end-states in different environments, and that they do so quite fast in dense star-forming gas \citep{Peters2017,Zhukovska2018}; this evolution is likely to alter $\beta$. Similarly, $T\subD$ is not expected to be uniform along the line of sight, because the radiation field that heats the dust is not uniform.

Finally, the standard procedure assumes that the dust emission is optically thin at all the observed wavelengths, and so the monochromatic intensity at wavelength $\lambda$ is
\begin{eqnarray}
I_{_\lambda}&\simeq&\tau_{_\lambda}\;\,B_{_\lambda}\!\!\left(T\subD\right)\;\,=\;\,\tau\rwl\left(\!\frac{\lambda}{\rm 300\,\si{\micro} m}\!\right)^{\!-\beta}\;\,B_{_\lambda}\!\!\left(T\subD\right).\hspace{0.8cm} \label{EQN:standInt}
\end{eqnarray}
Given a good signal in at least three distinct wavebands, there is in principle sufficient information to solve for $\tau\rwl$, $\beta$ and $T\subD$. In practice this works best if the wavebands are distributed in wavelength so that they sample emission from both well above, and well below, the peak of the spectrum. Since most of the dust in the L1495 Main Filament is in the temperature range $9\,{\rm K}\la T\subD\la 18\,{\rm K}$, and since we anticipate $\beta\la 2.0$, this requires a waveband with mean wavelength ${\bar\lambda}\la\,$\SI{70}{\micro\meter} and a waveband with mean wavelength ${\bar\lambda}\ga\,$\SI{1000}{\micro\meter}. The longest {\em Herschel} waveband has ${\bar\lambda}\simeq\,$\SI{500}{\micro\meter}, and so the long-wavelength side of the spectrum (the modified Rayleigh-Jeans tail) is not properly sampled. Consequently, there is a degeneracy, whereby high $T\subD$, can be mimicked by low $\beta$, and vice versa. To mitigate this problem, some authors fix $\beta\!=\!2.0$ \citep[since this is the value predicted by many theoretical grain models, e.g.][]{Mathis1990,Li2001,Draine2003} and simply solve for $\tau\rwl$ and $T\subD$.

%%%%%
\begin{figure*}
\includegraphics[width=\textwidth]{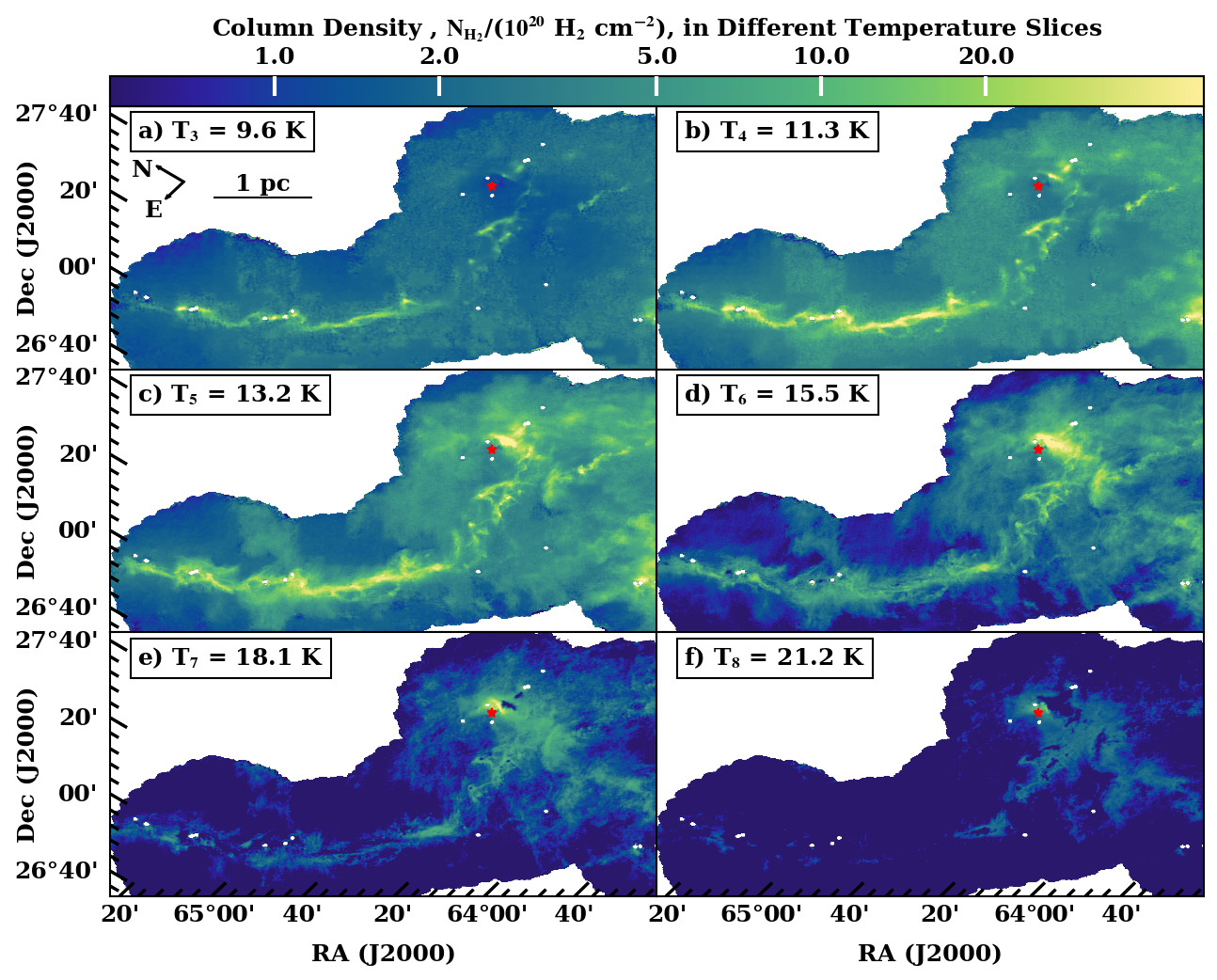}
\caption{Six contiguous temperature slices for the L1495 region. Each panel gives the distribution of dust with temperature close to the value marked in the top left corner. The white circles are masked-out, optically thick cores. The angular resolution is 18\si{\arcsecond}, which is equivalent to $\sim\!0.013\,{\rm pc}$ at the distance of Taurus. The colour-bar gives the column-density of molecular hydrogen, $N_{_{\rm H_2}}\!$, in units of $10^{20}\,{\rm H_2\,cm^{-2}}$. The red star indicates the position of V892 Tau, a Herbig Ae star. See text for further details. RA and Dec are marked by slanted ticks.}
\label{FIG:dNdT}
\end{figure*}
%%%%%

%%%%%
\begin{figure*}
\includegraphics[width=\textwidth]{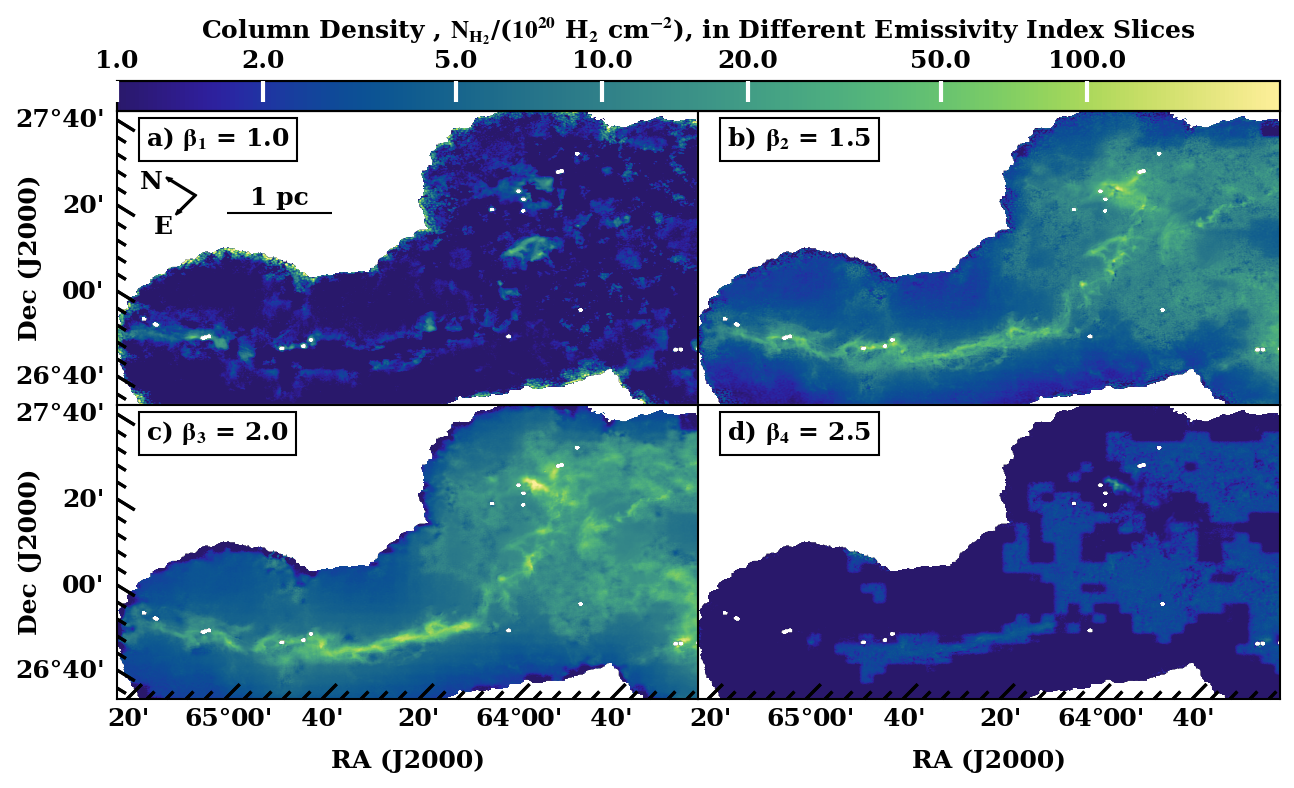}
\caption{Emissivity-index slices for the L1495 region. Each panel gives the distribution of dust with emissivity index close to the value marked in the top left corner. Other details are as in Fig. \ref{FIG:dNdT}.}
\label{FIG:dNdbeta}
\end{figure*}

%%%%%
\begin{figure*}
\includegraphics[width=\textwidth]{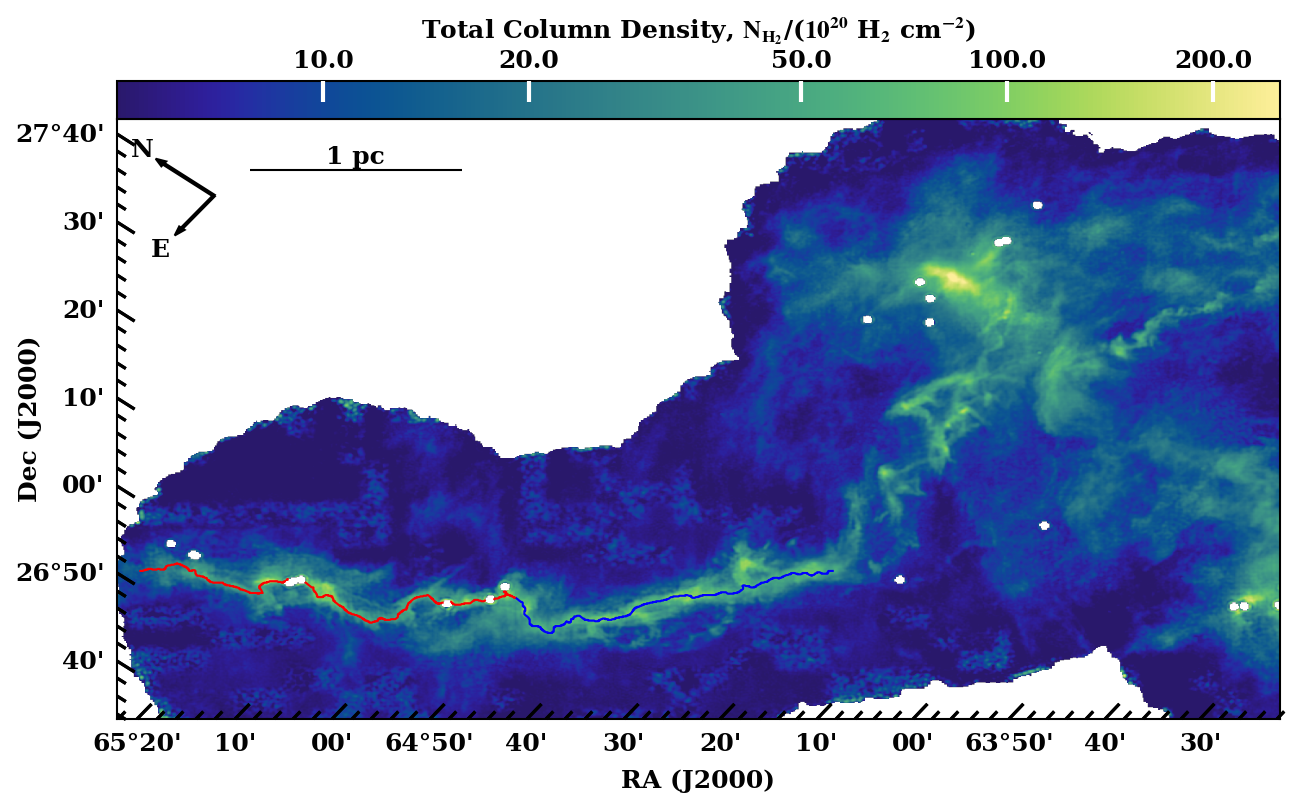}
\caption{Map of the total column density of molecular hydrogen, $N_{_{\rm H_2}}$ (derived from $\tau\rwl$ using eqn. \ref{EQN:tau2NH2}) for the L1495 region. The white circles are masked-out, optically thick cores. The red and blue lines delineate the spines of the B213 and B211 sub-filaments (which together make up the L1495 Main Filament), as identified by the DisPerSE algorithm (see Section \ref{SEC:cyl}). The region to the west of this is the L1495 Head. This paper is solely concerned with the L1495 Main Filament; a second paper will discuss the L1495 Head.}
\label{FIG:tauO}
\end{figure*}
%%%%%

%%%%%
\section{The {\sc PPMAP} procedure for analysing maps of thermal dust emission}\label{SEC:PPMAP}
%%%%%

Unlike the standard procedure, \ppp uses the input maps at their native resolution, and thereby retains the extra information contained in the maps that have finer resolution. The \ppp data products derived here have 18\si{\arcsecond} angular resolution, in order that we can make meaningful comparisons with the results obtained by \citetalias{Palmeirim2013} using the standard analysis procedure and the same angular resolution. The distance to Taurus is $D\!\simeq\!140\,{\rm pc}$ \citep{Elias1978}, so 18\si{\arcsecond} resolution corresponds to $\sim$\SI{0.012}{pc} (or $\sim$\SI{2500}{AU}).

Unlike the standard procedure, \ppp presumes that there will be different types of dust and different dust temperatures on the line of sight viewed by each map pixel. Different types of dust are represented by different discrete values of $\beta$, labelled $\beta_{_k}$. Here we use four linearly spaced values, $\,\beta_{_1}\!=\!1.0,\;\beta_{_2}\!=\!1.5,\;\beta_{_3}\!=\!2.0$ and $\beta_{_4}\!=\!2.5$. $\,\;\beta_{_1}\!=\!1.0$ is intended to represent emissivity indices in a small interval about 1.0, i.e. $0.75\la\beta\la 1.25$, and similarly for the other $\beta_{_k}$ values. Different dust temperatures are represented by different discrete values, labelled $T_{_\ell}$. Here we use twelve logarithmically spaced values, $\,T_{_1}\!=\!7.0\,{\rm K},\;T_{_2}\!=\!8.2\,{\rm K},\;T_{_3}\!=\!9.6\,{\rm K},\;T_{_4}\!=\!11.3\,{\rm K},\;T_{_5}\!=\!13.2\,{\rm K},\;T_{_6}\!=\!15.5\,{\rm K},\;T_{_7}\!=\!18.1\,{\rm K},\;T_{_8}\!=\!21.2\,{\rm K},\;T_{_9}\!=\!24.9\,{\rm K},\;T_{_{10}}\!=\!29.1\,{\rm K},\;T_{_{11}}\!=\!34.1\,{\rm K}$ and $\,T_{_{12}}\!=\!40.0\,{\rm K}.\;$ Again, $\,T_{_1}\!=\!7.0\,{\rm K}$ is intended to represent dust temperatures in a small range about $7.0\,{\rm K}$, i.e. $6.5\,{\rm K}\la T\subD\la 7.6\,{\rm K}$, and similarly for the other $T_{_\ell}$ values.

Like the standard procedure, the \ppp procedure assumes that the dust emission is optically thin. Any regions that are optically thick -- which in the context of star formation regions like the L1495 Main Filament means protostellar cores  \citep{Ossenkopf1994} -- must be ignored. To achieve this, bright peaks are located on the {\em Herschel}~PACS \SIadj{70}{\micro\meter} map, using the FellWalker algorithm \citep{Berry2015}, and those that correspond to point sources are identified and masked out using a circular patch with an angular diameter of \SI{120}{\arcsec} ($\sim 0.08\,{\rm pc}$); they appear as white dots on the maps. The masked sources correspond to dense, protostellar cores identified in \citet{Marsh2016}, and to the Class 0/I objects identified in \citet{Rebull2010}.

The intensity in each pixel, $(i,j)$, is then given by
\begin{eqnarray}
I_{_\lambda}\!&\!=\!&\!\sum\limits_{k=1}^{k=4}\,\sum\limits_{\ell=1}^{\ell=12}\,\left\{\Delta^2\tau_{_{300:k\ell}}\;\left(\!\frac{\lambda}{\rm 300\,\si{\micro} m}\!\right)^{\!-\beta_{_k}}\;B_{_\lambda}\!\!\left(T_{_\ell}\right)\right\},\hspace{0.8cm} \label{EQN:PPInt}
\end{eqnarray}
where $\Delta^2\tau_{_{300:k\ell}}$ is the contribution to the total optical depth at the reference wavelength,  $\tau\rwl$, from dust with $\beta\sim\beta_{_k}$ and $T\subD\sim T_{_\ell}$. Thus the raw data products from \ppp are four-dimensional data-cubes, with two dimensions representing position on the sky, $(x_{_i},y_{_j})$, one dimension representing the emissivity index, $\beta_{_k}$, and one dimension representing the dust temperature, $T_{_\ell}$. There are two data-cubes, one giving the expectation values for $\Delta^2\tau_{_{300:k\ell}}$, and the other giving the corresponding uncertainties $\Delta^2\unc_{_{300:k\ell}}$

\ppp generates these data-cubes using a Bayesian fitting algorithm. The algorithm starts by populating the data-cube with a uniform array of very small optical-depth quanta, $\delta\tau\rwl$, and then generates the maps that this configuration would produce in the different wavebands, with their different point spread functions. These synthetic maps are then compared with the real maps, assuming an extremely high level of synthetic noise, and the distribution of optical-depth quanta is adjusted, to produce a slightly better fit. Because the noise is high, the adjustments are small, i.e. in the linear regime. This process is performed iteratively, and at each iteration the synthetic noise is reduced, until it is completely removed. Details of the algorithm are given in \citet{Marsh2015}, along with a range of tests on synthetic data. The version described there is three-dimensional, with two dimensions representing position on the sky and one representing dust temperature, $T\subD$; the emissivity index is held constant at $\beta\!=\!2$. Extension to four dimensions, i.e. the introduction of different $\beta$ values, is mathematically trivial, but requires more computation. The algorithm invokes a tight prior on $\beta$, specifically a  Gaussian with mean $\mu_{_\beta}\!=\!2.0$ and standard deviation $\varSigma_{_{\!\beta}}\!=\!0.25$. This is necessary to regulate the $(\beta,T\subD)$ degeneracy, whereby -- given the limited wavelength range of the data \citep[e.g.][]{Shetty2009a,Shetty2009b} -- low $\beta$ can be mimicked by high $T\subD$, and {\it vice versa}. The tight prior on $\beta$ ensures that the algorithm only deviates from the canonical value of $\beta\!=\!2.0$ when the data really require this. A flat prior is used for $\log\left(T\subD\right)$.

%%%%%
\begin{figure*}
\centering
\includegraphics[width=0.9\linewidth]{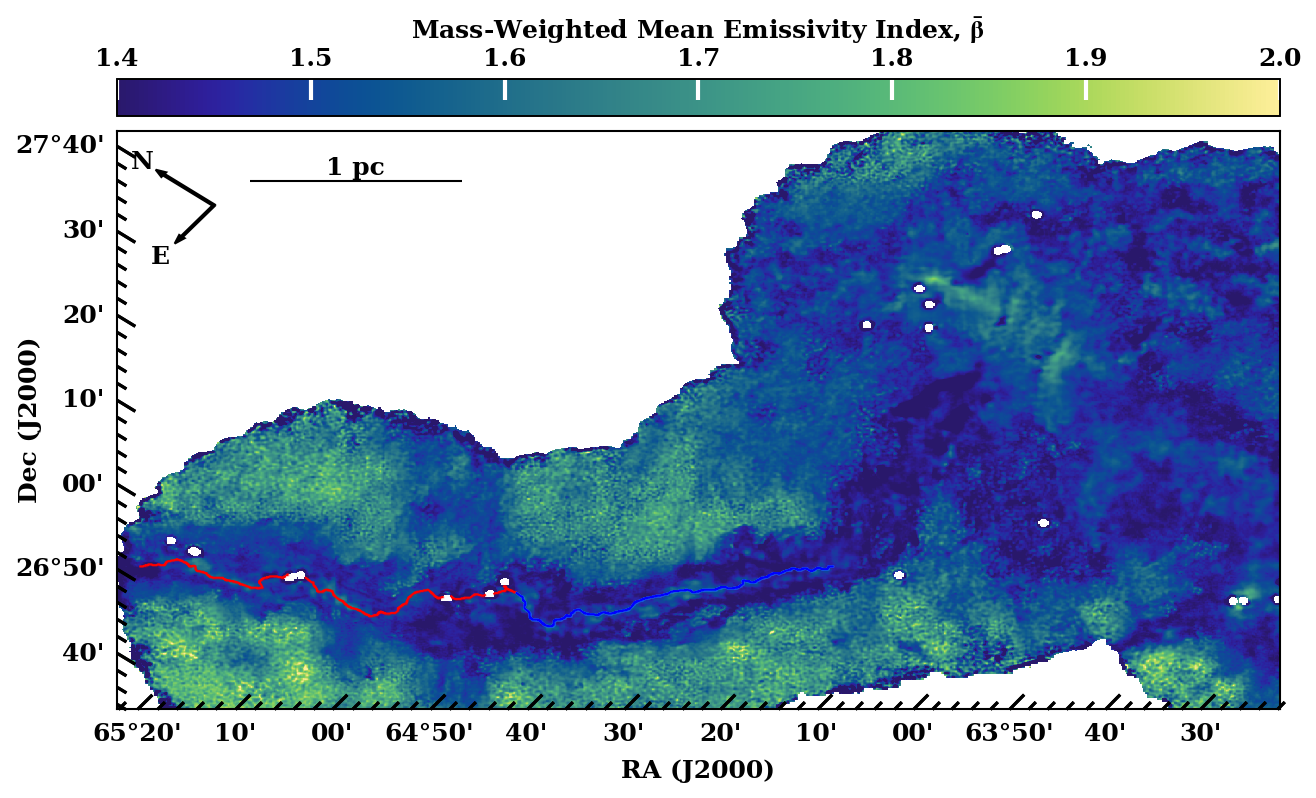}
\includegraphics[width=0.9\linewidth]{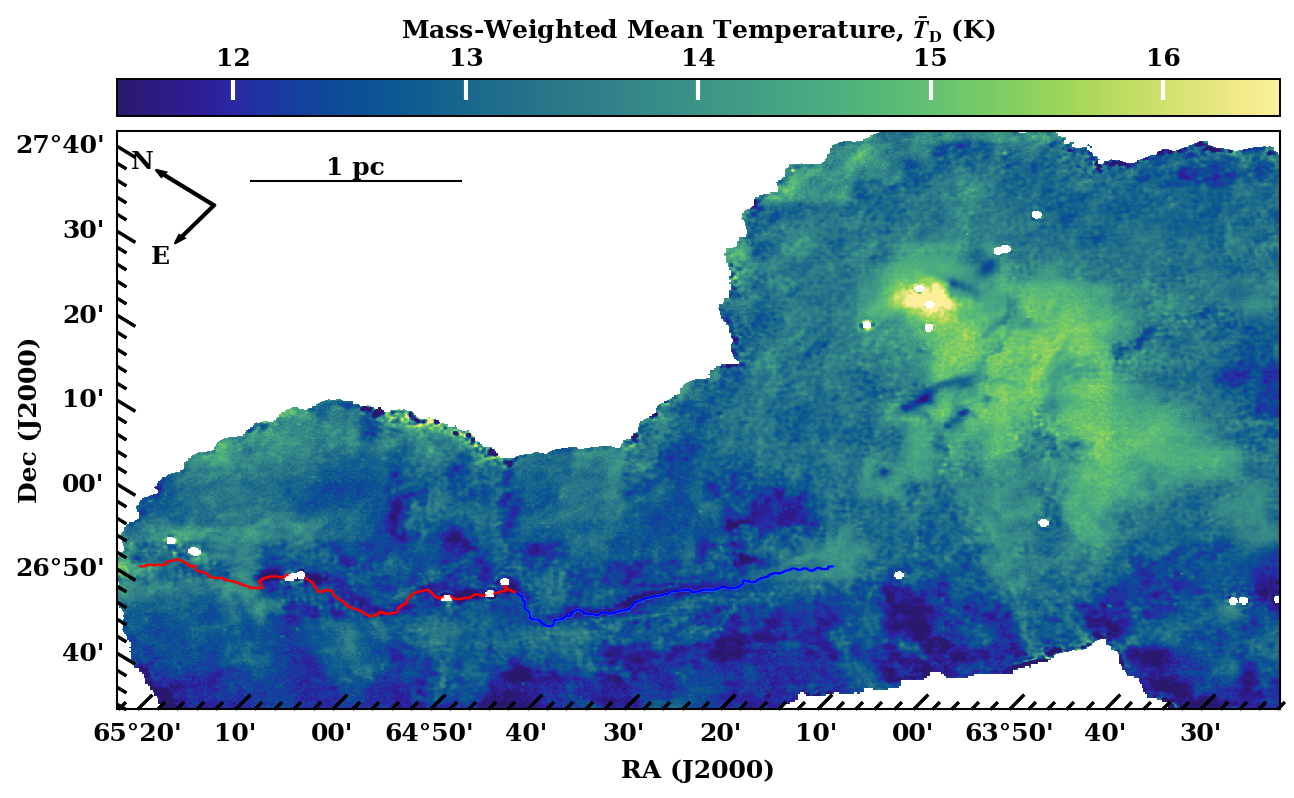}
\caption{The mean line-of-sight emissivity index, ${\bar\beta}$ (top panel) and dust temperature ${\bar T}\subD$ (bottom panel) for the  L1495 region. The white circles are masked-out, optically thick protostellar cores. The dust in the interior of the filament has lower emissivity index, and is cooler, than the dust in the outer layers and in the surroundings. The filament spines are shown as in Fig. \ref{FIG:tauO}.} 
\label{FIG:betabar;Tbar}
\end{figure*}
%%%%%

%%%%%
\section{Basic {\sc PPMAP} data products for L1495}\label{SEC:prods}
%%%%%

The full information contained in the raw four-dimensional \ppp data-cubes is hard to visualise. However, given the values of $\Delta^2\tau_{_{300:k\ell}}$, we can marginalise out one of the dimensions, to obtain a three-dimensional data-cube. For example, if we marginalise out $\beta$ (i.e. sum $\Delta^2\tau_{_{300:k\ell}}$ over all the discrete values, $\beta_{_k}$), we obtain
\begin{eqnarray}\label{EQN:Deltatau_OTL}
\Delta\tau_{_{300:\ell}}&=&\sum\limits_{k=1}^{k=4}\left\{\Delta^{\!2}\tau_{_{300:k\ell}}\right\}.
\end{eqnarray}
$\Delta\tau_{_{300:\ell}}$ is the contribution to $\tau\rwl$, from dust at temperature $T_{_\ell}$ (or, strictly speaking, from dust in the small temperature interval represented by $T_{_\ell}$). Maps of $\Delta\tau_{_{300:\ell}}(x,y)$ are analogous to the position-position-velocity slices derived from spectral-line observations, but with velocity replaced by dust temperature, and integrated intensity replaced by optical depth. We will refer to them as `temperature slices'.

Fig. \ref{FIG:dNdT} shows temperature slices for the L1495 Main Filament and surroundings, at six contiguous temperatures ($\,T_{_3}\!=\!9.6\,{\rm K},\;T_{_4}\!=\!11.3\,{\rm K},\;T_{_5}\!=\!13.2\,{\rm K},\;T_{_6}\!=\!15.5\,{\rm K},\;T_{_7}\!=\!18.1\,{\rm K},\;T_{_8}\!=\!21.2\,{\rm K}$). For ease of interpretation, the colour bar gives the corresponding column-density of molecular hydrogen, which is obtained by multiplying $\Delta\tau_{_{300:\ell}}$ by $2.1\times 10^{24}\;{\rm H}_{_2}\,{\rm cm}^{-2}$ (see Eqn. \ref{EQN:tau2NH2}). However, we should be mindful that what is traced here -- and in other maps -- is dust. These slices show that the cold dust ($T\subD\la 15\,{\rm K}$) is concentrated in the filament, with the coldest dust close to the filament spine, whilst the warmer dust ($T\subD\ga 17\,{\rm K}$) is distributed throughout the surroundings. There is an area of especially warm dust in the vicinity of the Herbig Ae star V892 Tau, at $\mbox{RA}=64^{\rm o}\,40'\,15'',\;\mbox{Dec}=28^{\rm o}\,19'\,16''$. This is presumably due to extra local heating from this energetic star, which is known to be producing X-ray flares \citep{Guardino2004}. The position of V892 Tau is marked with a red star on Fig.~\ref{FIG:dNdT}.

Similarly, if we marginalise out $T\subD$ (i.e. sum $\Delta^2\tau_{_{300:k\ell}}$ over all the discrete values, $T_{_\ell}$), we obtain
\begin{eqnarray}\label{EQN:Deltatau_OBK}
\Delta\tau_{_{300:k}}&=&\sum\limits_{\ell=1}^{\ell=12}\left\{\Delta^{\!2}\tau_{_{300:k\ell}}\right\}.
\end{eqnarray}
$\Delta\tau_{_{300:k}}$ is the contribution to $\tau\rwl$ from dust with emissivity index $\beta_{_k}$ (strictly speaking, from dust in the small range of emissivity index represented by $\beta_{_k}$). We will refer to maps of $\Delta\tau_{_{300:k}}(x,y)$ as `emissivity-index slices'.

Fig. \ref{FIG:dNdbeta} shows emissivity-index slices for the L1495 Main Filament and surroundings, at all four $\beta$ values. Again, the colour bar gives the corresponding column-density of molecular hydrogen, based on the conversion factor in Eqn. (\ref{EQN:tau2NH2}). These slices show that the dust in the surroundings and the outer filament sheath has $\beta\ga 1.7$, while the dust near the spine of the filament has $\beta\la 1.5$, and a few small dense regions even exhibit values of $\beta \sim 1.0$.

If we marginalise out both $\beta$ and $T\subD$, we obtain the total optical depth at the reference wavelength,
\begin{eqnarray}\label{EQN:tau_O}
\tau\rwl&=&\sum\limits_{\ell =1}^{\ell =12}\,\sum\limits_{k =1}^{k =4}\left\{\Delta^{\!2}\tau_{_{300:k\ell}}\right\}\,.
\end{eqnarray}
The corresponding total uncertainty map is obtained by adding the individual contributions from different combinations of $\beta_{_k}$ and $T_{_\ell}$ in quadrature, i.e.
\begin{eqnarray}\label{EQN:unc_O}
\unc^2\rwl&=&\left(\sum\limits_{\ell =1}^{\ell =12}\,\sum\limits_{k =1}^{k =4}\left\{\left(\Delta^{\!2}\unc_{_{300:k\ell}}\right)^2\right\}\right)^{1/2}\,.
\end{eqnarray}

Fig. \ref{FIG:tauO} shows a map of $\tau\rwl$ for the L1495 region. Again, the scale bar gives the corresponding column-density of molecular hydrogen, based on the conversion factor in Eqn. (\ref{EQN:tau2NH2}). The white circles are the masked-out, optically thick cores. The region to the west of the break at $\mbox{\sc RA}\sim64^{\rm o} 28'$ and $\mbox{\sc Dec}\sim27^{\rm o} 22'$ will be referred to as the L1495 Head. The very elongated region to the east of this break will be referred to as the L1495 Main Filament, and this is the region with which this paper is concerned. The red and blue lines mark the spines of, respectively, the B213 and B211 sub-filaments, which together comprise the L1495 Main Filament (see Section \ref{SEC:cyl} for details of how the spine is located).

For each pixel, $(i,j)$, we can define the line-of-sight mean emissivity index, 
\begin{eqnarray}\label{EQN:beta_bar}
{\bar\beta}&=&\frac{1}{\tau\rwl}\;\;\sum\limits_{\ell =1}^{\ell =12}\,\sum\limits_{k =1}^{k =4}\left\{\beta_{_k}\;\,\Delta^{\!2}\tau_{_{300:k\ell}}\right\}\,,
\end{eqnarray}
and the line-of-sight mean dust temperature,
\begin{eqnarray}\label{EQN:T_bar}
{\bar T}\subD&=&\frac{1}{\tau\rwl}\;\;\sum\limits_{\ell =1}^{\ell =12}\,\sum\limits_{k =1}^{k =4}\left\{T_{_\ell}\;\,\Delta^{\!2}\tau_{_{300:k\ell}}\right\}\,.
 \end{eqnarray}

Fig. \ref{FIG:betabar;Tbar} shows maps of ${\bar\beta}$ (top panel) and ${\bar T}\subD$ (bottom panel) in the L1495 region. ${\bar\beta}$ is clearly significantly lower in the filament, ${\bar\beta}\la 1.5$, than in the surroundings, ${\bar\beta}\ga 1.7$, suggesting that physical conditions in the filament have effected a change in the properties of the dust. This might be a consequence of the increased density in the filament, promoting dust growth. However, given the width of the region with reduced ${\bar\beta}$, it might also be due to processes in the accretion shock where the material that is now in the filament flowed onto the filament. Evidence for shocks is provided by \citetalias{Palmeirim2013}, where accreting material is estimated to have an inflow velocity of between \SIlist{0.5;1.0}{\kilo \meter \per \second} compared to a sound speed of \SI{0.19}{\kilo \meter \per \second} (and thus a Mach Number between 2.6 and 5.3). In contrast, the temperature shows a much narrower minimum, near the spine of the filament, which we attribute to attenuation of the ambient radiation field. The temperature is also quite low in the background to the south side of the filament, and this suggests that the ambient radiation field on the south side is somewhat weaker than on the north side; assuming $T\subD\propto U_{_{\rm RAD}}^{1/(4+p)}$, where $U_{_{\rm RAD}}$ is the ambient radiation density, and $p\simeq 2$, a factor $U_{_{\rm RAD.South}}\sim 0.67U_{_{\rm RAD.North}}$ would suffice.

%%%%%
\begin{figure}
\centering
\includegraphics[width=0.8\linewidth]{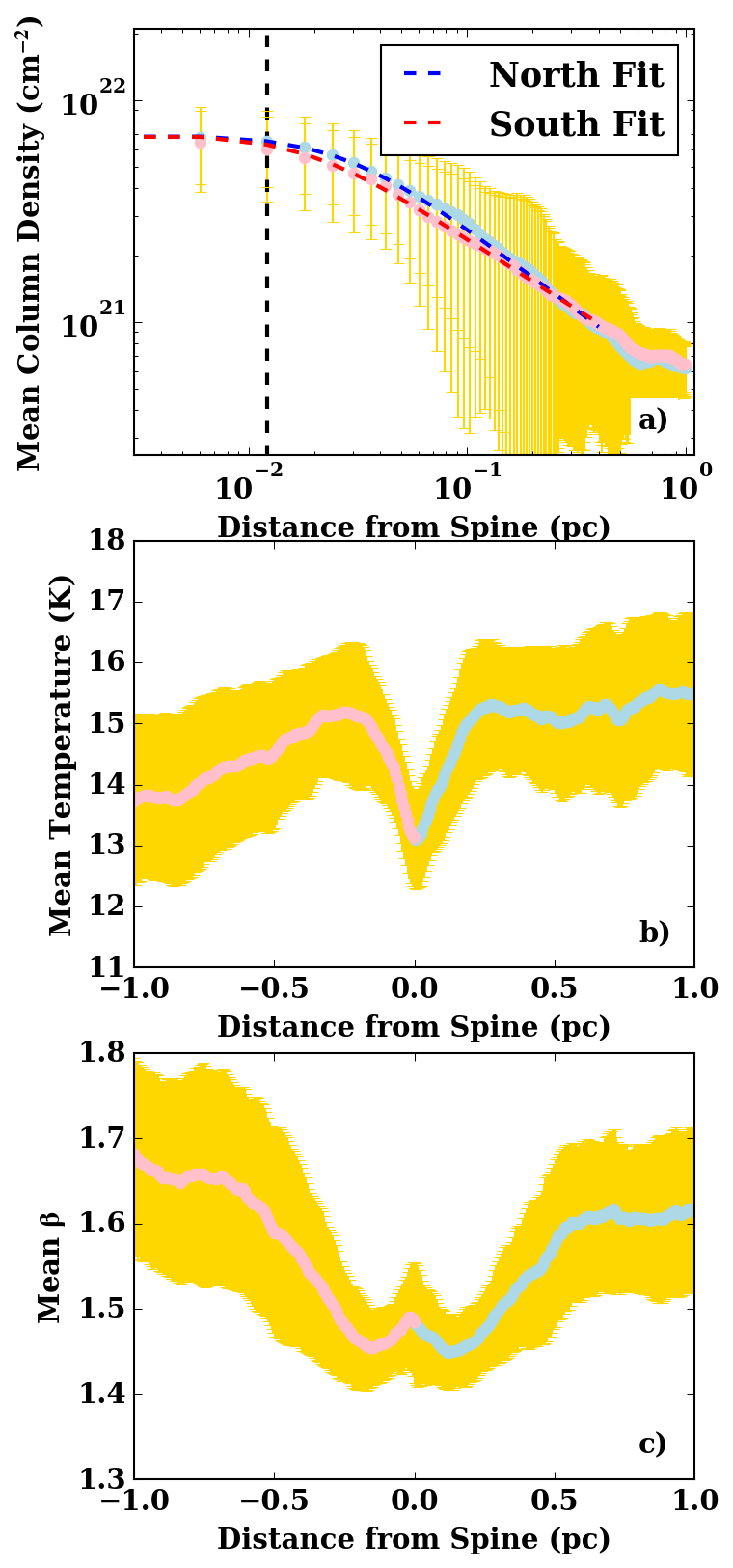}
\caption{Profiles of (a) the column density, $N_{_{\rm H_2}}\!(b)$, (b) the line-of-sight mean temperature, ${\bar T}\subD(b)$, and (c) the line-of-sight mean emissivity index, ${\bar\beta}(b)$, all three as functions of impact parameter, $b$ (measured relative to the spine of the filament). Pale blue [pink] dots represent mean values from the north [south] side of the filament, obtained by averaging the values at discrete $b$ values from all 1032 {\it Local Sample Profiles}; at large $b$ the dots merge into a continuous curve. The yellow shading represents the standard deviation on these mean values. On (a) the blue [red] dashed lines give Plummer-like fits to the mean column-density on the north [south] side (Eqn. \ref{EQN:NH2}, with parameters as given in Table \ref{TAB:fit}); the vertical dashed line indicates the \SI{18}{\arcsecond} resolution.The rise in the value of ${\bar\beta}$ near the spine in (c) is barely significant statistically, and we do not have a physical explanation for it; it may be a manifestation of the $(\beta ,T\subD)$ degeneracy.}
\label{FIG:profiles}
\end{figure}
%%%%%

%%%%%
\section{Analysis of the  L1495 Main Filament assuming cylindrical symmetry}\label{SEC:cyl}
%%%%%

In this section we analyse the \ppp results for the L1495 Main Filament, on the assumption that locally its cross-section is cylindrically symmetric. We consider measures of internal substructure within the filament in Section \ref{SEC:sub}.

First, we identify the spine of the L1495 Main Filament (i.e. the projection on the sky of the putative local axis of cylindrical symmetry) by applying the DisPerSE algorithm \citep{Sousbie2011} to the total optical-depth map, converted into column density using Eqn. (\ref{EQN:tau2NH2}). We invoke a persistence threshold of $10 \times 10^{20}\,{\rm H_2\,cm^{-2}}$. Then we use the {\sc skelconv}~algorithm, with a smoothing length of 5 pixels, to smooth the returned spines, and to trim spines with values below $12\times 10^{20}\,{\rm H_2\,cm^{-2}}$. In order to combine smaller spines into larger ones, the \verb|assemble| option is enabled, with an acceptance angle of \SI{60}{\degree}. Where this is deemed appropriate, the resultant spines are also joined manually, to produce a single continuous spine. The resulting spine is delineated by the red and blue lines on Fig. \ref{FIG:tauO}.

Next, we define 1100 discrete {\it Sample Points} along the spine, equally spaced at intervals of $\sim 0.004\,{\rm pc}$, and at each {\it Sample Point} we determine the local tangent to the spine by spline fitting.

In what follows, we use the variable $r$ for the true 3D distance from the spine (e.g. Eqn. \ref{EQN:rhor} below), and the variable $b$ for projected 2D distance from the spine (i.e. the impact parameter of the line of sight; e.g. Eqn. \ref{EQN:NH2} below). It is important to be mindful of the distinction between these two different distances.

%%%%%
\subsection{The mean profiles of L1495}\label{SEC:MeanProfile}
%%%%%

Inside the filament, we assume that the volume-density of molecular hydrogen, $n_{_{\rm H_2}}\!(r)$, subscribes to a Plummer-like profile \citep[cf.][]{Whitworth2001,Nutter2008,Arzoumanian2011},
\begin{eqnarray}\label{EQN:rhor}
n_{_{\rm H_2}}(r)&=&n\subO\,\left\{1+\left(\frac{r}{r\subO}\right)^2\right\}^{-p/2}\!,\hspace{0.8cm}r<r\subB\,.\hspace{0.8cm}
\end{eqnarray}
Here, $n\subO$ is the volume-density of molecular hydrogen on the spine, and $r\subO$ is the radius within which the volume-density is approximately uniform; $p$ is the asymptotic radial density exponent, i.e. $p=-\,d\ln\left(n_{_{\rm H_2}}\right)/d\ln(r)$ for $r\gg r\subO$; $r\subB$ is the boundary of the filament, outside of which the volume-density is presumed to be approximately uniform.\footnote{The density profile of an equilibrium self-gravitating isothermal filament is given by Eqn. (\ref{EQN:rhor}) with $p\!=\!4$ and $r\subO=a\subO(2/\pi Gn\subO{\bar m}_{_{\rm H_2}})^{1/2}$, where $a\subO$ is the isothermal sound speed \citep{Ostriker1964}.}

Given this volume-density profile, and ignoring curvature of the filament, the column-density of molecular hydrogen, $N_{_{\rm H_2}}\!(b)$ (as displayed in Fig. \ref{FIG:tauO}), can be fit with
\begin{eqnarray}\label{EQN:NH2}
N_{_{\rm H_2}}\!(b)&=&N\subO\left\{1+\left(\frac{b}{r\subO}\right)^2\right\}^{-(p-1)/2}\;\,+\;\,N\subB\,,\\\label{EQN:NO}
N\subO&=&n\subO\,r\subO\;\,B\!\left(\frac{1}{2},\frac{(p-1)}{2}\right)\;\,\sec(i)\,.
\end{eqnarray}
Here $N\subO$ is the excess column-density through the spine, and $N\subB$ is the background column-density; $B$ is the Euler Beta Function \citep{Casali1986}; $i$ is the inclination of the filament to the plane of the sky. 

For each of the 1100 {\it Sample Points} along the spine of the filament, we know the local tangent. The {\it Local Sample Profile} is determined by computing the column-density at discrete impact parameters $\pm j\Delta b$ along a cut through the corresponding {\it Sample Point} and orthogonal to the local tangent. Here `$+$' (`$-$') refers to displacements to the north (south) side of the spine; $j$ is an integer on the interval $[0,164]$; and $\Delta b=0.006\,{\rm pc}$. The {\it Local Sample Profile} is therefore defined by $1+2\!\times\!164 =  329$ column-densities, and extends out to lines of sight displaced $\sim 1\,{\rm pc}$ from the spine of the filament.

For 68 {\it Sample Points} the {\it Local Sample Profiles} are corrupted by poor data or masked-out protostars. We combine the remaining 1032 {\it Local Sample Profiles} by taking -- at each of the 329 discrete impact parameters, $\pm j\Delta b$ -- the median of the 1032 column-densities on the 1032 {\it Local Sample Profiles}, to obtain a single {\it Global Average Profile} (comprising 329 column-densities). We then fit this {\it Global Average Profile} with Eqn. (\ref{EQN:NH2}) using the \verb|LMFIT| Python package \citep{Newville2014}. Since $N\subB$ appears to be different on the two sides of the filament, we fit the north side separately from the south side, and obtain seven fitting parameters: $N\subO$, $(r\subO,p,N\subB)_{_{\rm North}}$ and $(r\subO,p,N\subB)_{_{\rm South}}$. Given $N\subO$, $r\subO$ and $p$, Eqn. (\ref{EQN:NO}) can be inverted to obtain $n\subO\sec(i)$, and hence an upper limit on $n\subO$.

Table \ref{TAB:fit} gives these fitting parameters and their uncertainties. Fig. \ref{FIG:profiles}(a) shows the {\it Global Average Profile}, $\bar{N}_{_{\rm H_2}}\!(b)$, and the associated standard deviation. Relative to the north profile, the south profile has smaller scale-length, $r\subO$, smaller density exponent, $p$, and smaller background column-density, $N\subB$. This may indicate that the inflow onto the filament from the south delivers a slightly higher ram-pressure than that from the north, but the difference is not great.

Figs. \ref{FIG:profiles}(b) and (c) show the mean line-of-sight temperature profile, ${\bar T}\subD (b)$, and the mean line-of-sight emissivity index profile, ${\bar\beta}(b)$. Both ${\bar T}\subD$ and ${\bar\beta}$ tend to decrease towards the spine of the filament. However, as already discussed in Section \ref{SEC:prods}, the ${\bar T}\subD$ minimum is much narrower than the ${\bar\beta}$ minimum, so ${\bar T}\subD$ and ${\bar\beta}$ are not significantly correlated. We note that, because ${\bar\beta}$ and $\bar{T}\subD$ are line-of-sight means, their dynamic range is not a faithful indicator of the full dynamic range of $\beta$ and $T\subD$.

%%%%%
\begin{table}
\centering
\caption{Mean values and standard deviations for the parameters derived by fitting the column density of molecular hydrogen, $N_{_{\rm H_2}}\!(b)$ with a Plummer-like profile (Eqn. \ref{EQN:NH2}); here $b$ is the impact parameter of the line of sight, relative to the spine of the filament, and the north and south sides of the filament are fit separately. The first four parameters ($N\subO$, $r\subO$, $p$, $N\subB$) are varied to obtain the fit, and the remaining parameter ($n\subO$) is derived from the fit. If the inclination of the filament to the plane of the sky is $i$, $N\subO$ and $n\subO$ should be multiplied by $\cos(i)$.}
\label{TAB:fit}
\begin{tabular}{rccc}\hline
{\sc Parameter} & \hspace{0.2cm} & \hspace{0.4cm}{\sc North}\hspace{0.4cm} & \hspace{0.4cm}{\sc South}\hspace{0.4cm} \\
$N\subO/(10^{21}\,{\rm H_2\,cm^{-2}})$ & & $6.38\pm 0.09$  & $6.38\pm 0.09$ \\
$r\subO/(10^{-2}\,{\rm pc})$ & & $2.80\pm 0.14$ & $1.67\pm 0.06$ \\
$p$ & & $2.02\pm 0.05$ & $1.73\pm 0.02$ \\
$N\subB/(10^{21}\,{\rm H_2\,cm^{-2}})$ & & $0.28\pm 0.03$ & $0.13\pm 0.04$ \\
$n\subO/(10^4\,{\rm H_2\,cm^{-3}})$ & & $2.39\pm 0.70$ & $3.15\pm 0.97$ \\\hline
\end{tabular}
\end{table}
%%%%%

%%%%%
\subsection{The mean global parameters of the L1495 Main Filament}\label{SEC:MeanGlobal}
%%%%%

From the {\it Global Average Profile}, the average column-density of molecular hydrogen through the spine of the filament is $\overline{N}\subO\simeq 6.4\times 10^{21}\,{\rm H_2\,cm^{-2}}$ (significantly less than the value, $\overline{N}\subO \simeq 16\times 10^{21}\,{\rm H_2\,cm^{-2}}$, obtained by \citetalias{Palmeirim2013}). If we adopt a Plummer-like profile with $\overline{p}\!=\!0.5(p_{_{\rm North}}+p_{_{\rm South}})=1.88$, $\overline{r}\subO=0.5\left(r_{_{\rm O:North}}+r_{_{\rm O:South}}\right)=0.022\,{\rm pc}$, $N\subO\simeq 6.4\times 10^{21}\,{\rm H_2\,cm^{-2}}$, and $r\subB =0.4\,{\rm pc}$, the mean {\sc fwhm} is
\begin{eqnarray}\label{EQN:FWHMPlummer}
\overline{\textsc{fwhm}}_{_{\rm \,Plummer}}&\simeq&0.087\pm 0.003\,{\rm pc}\,,
\end{eqnarray}
in close agreement with the  $\sim\!0.09\,{\rm pc}$ obtained by \citetalias{Palmeirim2013} and \citet{Arzoumanian2019}). However, the mean line-density is
\begin{eqnarray}\label{EQN:muPlummer}
\overline{\mu}_{_{\rm \,Plummer}}&=&26.8\,{\rm M}_{_\odot}\,{\rm pc}^{-1}\;\cos(i)\,,
\end{eqnarray}
significantly less than the $54\,{\rm M}_{_\odot}\,{\rm pc}^{-1}$ obtained by \citetalias{Palmeirim2013}.

These estimates (Eqns. \ref{EQN:FWHMPlummer} and \ref{EQN:muPlummer}) are the appropriate ones to compare with \citet{Arzoumanian2011} and \citetalias{Palmeirim2013}, but they are misleading on two counts, and will not be used in our subsequent analysis. First, the distributions of $N\subO$ and $r\subO$ for individual {\it Local Sample Profiles} are skewed towards high values, and therefore the {\sc fwhm} and line-density of the {\it Global Average Profile} do not accurately reflect the bulk of the filament; they are artificially inflated, as we demonstrate in Section \ref{SEC:LocFilProf}.

Second, if we were to follow the procedure used by \citet{Arzoumanian2011}, and also adopted by \citetalias{Palmeirim2013}, we would obtain a much smaller {\sc fwhm}. In their procedure, the centre of the Plummer-like profile is fit with a Gaussian, and the {\sc fwhm} of this Gaussian is adopted. The scale-length of a Gaussian fit to a Plummer-like profile is $\varSigma\subO =r\subO/(p-1)^{1/2}$, so we would have $\varSigma\subO\sim 0.024\,{\rm pc}$ and
\begin{eqnarray}
\overline{\textsc{fwhm}}_{_{\rm \,Gaussian}}\!&\!=\!&\! \left(8\ln(2)\right)^{1/2}\varSigma\subO\;\simeq\;0.056\pm 0.002\,{\rm pc}\,,\hspace{0.4cm}
\end{eqnarray}
significantly smaller than their estimate of $0.09\,{\rm pc}$. In other words, the apparent agreement between our {\sc fwhm} (based on a Plummer-like fit) and the {\sc fwhm} obtained by \citet{Arzoumanian2011} and \citetalias{Palmeirim2013} (based on a Gaussian fit) is fortuitous. In reality \ppp has enabled us to resolve the filament more accurately, and it is narrower. The mean line-density from a Gaussian fit is also significantly smaller, viz.
\begin{eqnarray}\label{EQN:muGaussian}
\overline{\mu}_{_{\rm \,Gaussian}}\!&\!\!=\!&(2\pi)^{1/2}N\subO{\bar m}_{_{\rm H_2}}\varSigma\subO\;\simeq\;8.6\,{\rm M}_{_\odot}\,{\rm pc}^{-1}\,\cos(i).\hspace{0.3cm}
\end{eqnarray}
The reason why $\overline{\textsc{fwhm}}_{_{\rm \,Gaussian}}\!<\!\overline{\textsc{fwhm}}_{_{\rm \,Plummer}}$ and ${\bar\mu}_{_{\rm \,Gaussian}}\!\ll\!{\bar\mu}_{_{\rm \,Plummer}}$ is that the Gaussian profile falls off at large radii much more rapidly than a Plummer-like profile with $p\sim 2$. This makes a significant difference at the half-maximum point, and an even bigger difference beyond this, in the outer layers of the filament, so the effect on the line-density is very large.

In what follows we will limit consideration to parameters obtained from Plummer-like fits to small local segments of the filament.

%%%%%
\begin{figure*}
\centering
\includegraphics[width=1.0\linewidth]{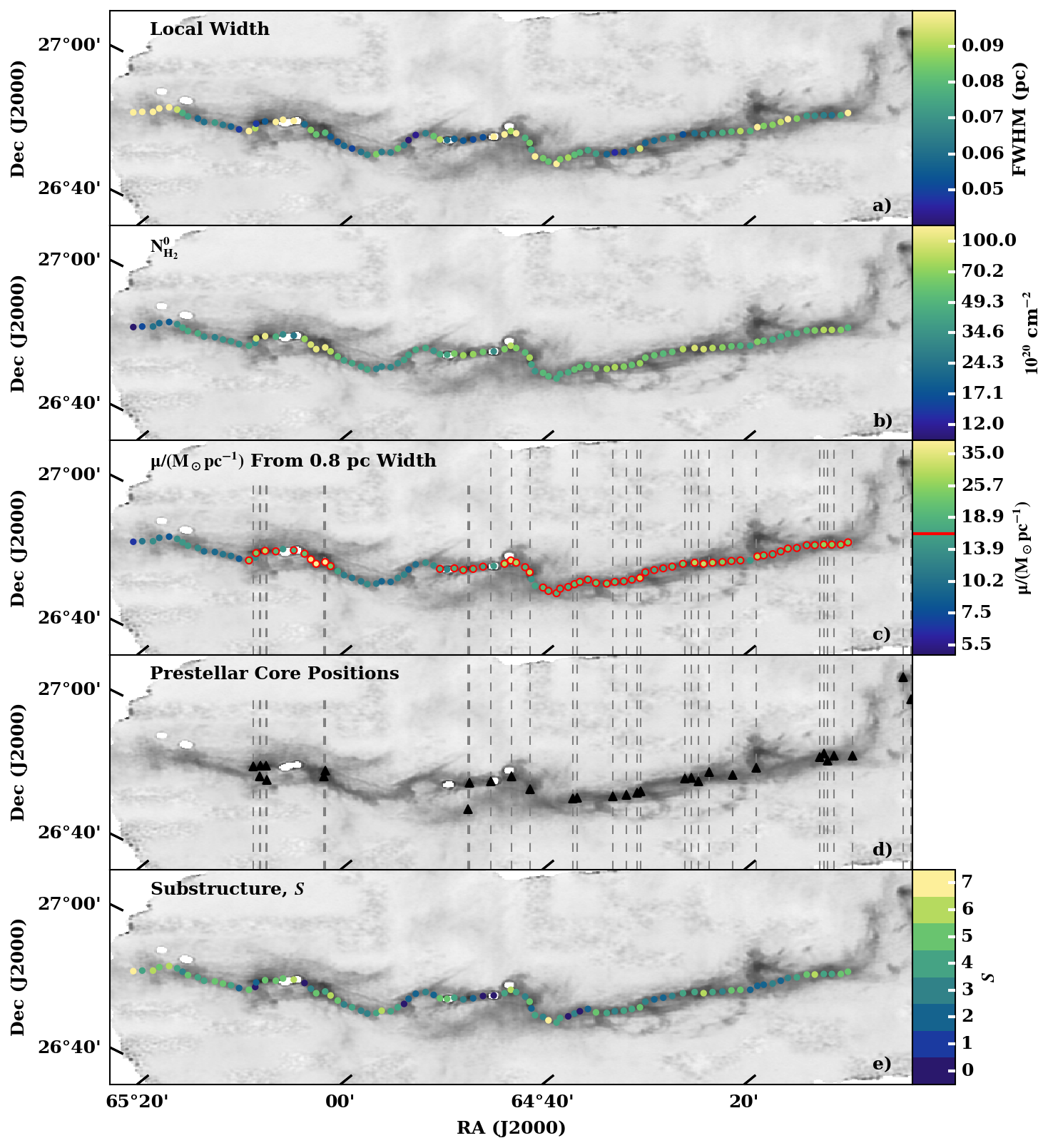}
\caption{Maps illustrating how the various global properties of the filament vary along its length. Each coloured circle sits on one of the $0.05\,{\rm pc}$-long segments created by bundling together 12 neighbouring sample points along the spine, and there are 92 segments in total. The colour-coding of the circles represents the local values of (a) the column-density of molecular hydrogen, $N\subO$, through the spine and with the background subtracted; (b) the local {\sc fwhm} of the filament; (c) the local line-density, $\mu$, of the filament (again, with the background subtracted); (d) the position of prestellar cores from \citet{Marsh2015}; and (e) the parameter ${\cal S}$, measuring of the level of resolved internal substructure within the filament (see Section \ref{SEC:sub}). The red horizontal line on the colour bar for (c) indicates the critical line-density, as given by Eqn.~\ref{EQN:muCRIT}, whilst red outlines around coloured circles represent regions that exceed this value. Vertical dashed grey lines in (c) and (d) show the horizontal positions of the prestellar cores.}
\label{FIG:SegmentProps}
\end{figure*}
%%%%%

%%%%%
\subsection{Variation along the L1495 Main Filament, setting \texorpdfstring{$\boldsymbol{p=2}$}{p=2}}\label{SEC:LocFilProf}
%%%%%

To evaluate variations along the length of the filament, we use the {\sc FilChaP}\footnote{\url{https://github.com/astrosuri/filchap}} algorithm \citep{Suri2018} to divide the filament into 92 contiguous {\it Segments}. Each {\it Segment} is constructed from 12 contiguous {\it Sample Points}, and is approximately \SI{\sim 0.05}{pc} (\SI{\sim72}{\arcsecond}) long. Thus the extent of a {\it Segment} along the filament is comparable to the filament {\sc fwhm}. For each {\it Segment}, we construct a {\it Segment Average Profile}, again by adopting -- at each of the 329 discrete impact parameters, $\pm j\Delta b$ -- the median of the column-densities from the 12 constituent {\it Local Sample Profiles}.

%%%%%
\begin{figure}
\centering
\includegraphics[width=1.0\linewidth]{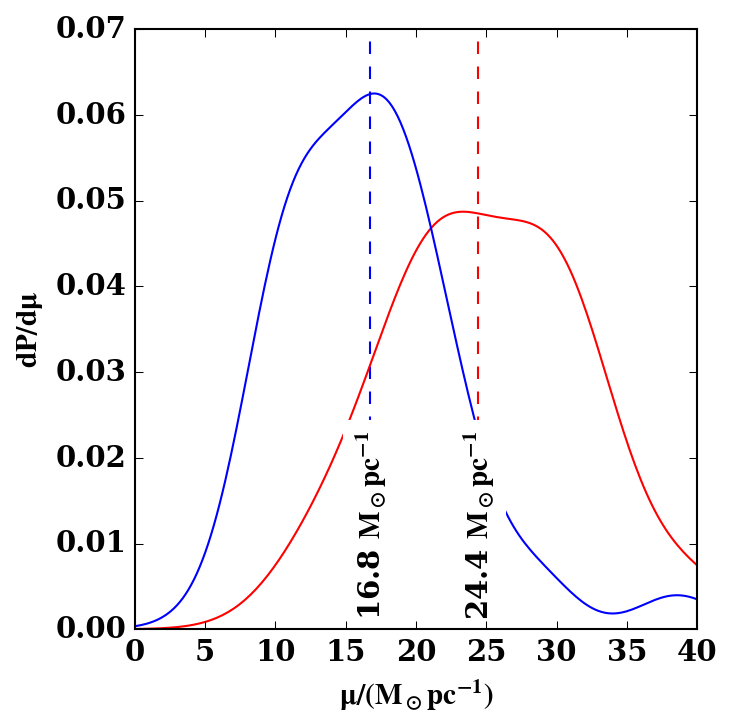}
\caption{Kernel-smoothed PDF of $\mu$ for the 20 segments that are closest to the 24 prestellar cores (red), and for the remaining 72 segments (blue). The medians are marked by vertical dashed lines.}
\label{fig:coreKDE}
\end{figure}
%%%%%

We then analyse each {\it Segment} independently, by fitting a Plummer-like profile (Eqn. \ref{EQN:NH2}) to the {\it Segment Average Profile}. The {\it Segment Average Profiles} are quite noisy, because each one is constructed from just 12 {\it Local Sample Profiles} (in contrast with the {\it Global Average Profile}, which is constructed from 1032 {\it Local Sample Profiles}). Therefore we do not attempt to solve for $p$, we simply set $p\!=\!2$.\footnote{This choice is somewhat arbitrary, and indeed a better fit can be obtained with $p\!=\!4$, as we show in Section \ref{SEC:p4exp}. However, it allows us to make straightforward comparisons with the earlier results of \citet{Arzoumanian2011} and \citetalias{Palmeirim2013}, and it has little influence on the global parameters ({\sc fwhm}, $\mu$) that we derive.} Since {\sc FilChaP} automatically performs a background subtraction, we fit each of the resulting {\it Local Sample Profiles} using Eqn. (\ref{EQN:NH2}) with $N\subB\!=0$ and $p\!=\!2$. This gives mean values of $N\subO$, and $r\subO$ for each {\it Segment}.

Fig.~\ref{FIG:SegmentProps}(a) shows how the estimated {\sc fwhm} (which for $p\!=\!2$ is given by $\textsc{fwhm}=12^{1/2}r\subO$)
varies along the filament. Each small circle marks the position of an individual {\it Segment}, and is colour coded to represent the estimated value of the {\sc fwhm}. Whilst the estimates at places where the filament sharply changes direction may be less well defined, the general picture is of a narrow filament. The median and interquartile range are $\textsc{fwhm} = 0.08^{+0.02}_{-0.02}\,{\rm pc}$.

Fig.~\ref{FIG:SegmentProps}(b) shows how the column-density of molecular hydrogen, $N\subO$, varies along the spine. The small circles are colour coded to represent the estimated value of $N\subO$ for that {\it Segment}. The median and interquartile range are $N\subO =4.93^{+2.80}_{-1.65}\times 10^{21}\,{\rm H_2\,cm^{-2}}$.

Fig.~\ref{FIG:SegmentProps}(c) shows how the line-density, $\mu$ (Eqn. \ref{EQN:muPlummer}), varies along the filament. The small circles are colour coded to represent the estimated value of $\mu$ for that {\it Segment}. The median and interquartile range are $\mu = 17.8^{+6.9}_{-6.6}\,{\rm M}_{_\odot}\,{\rm pc}^{-1}$. If we divide the filament into the B213 and B211 sub-filaments (marked, respectively, red and blue on Fig.~\ref{FIG:tauO}), we find that B211 has the larger median line-density, $\sim 19.9\,{\rm M}_{_\odot}\,{\rm pc}^{-1}$, and B213  the smaller, $\sim 13.4\,{\rm M}_{_\odot}\,{\rm pc}^{-1}$.

We do not have robust constraints on the isothermal sound speed, $c\subS$, in the L1495 Main Filament, but, if we assume that the gas is isothermal with $c\subS\!=\!0.19\,{\rm km\, s^{-1}}$ (corresponding to molecular gas with gas-kinetic temperature $T\!\simeq\!10\,{\rm K}$), the critical line-density \citep{Ostriker1964, Inutsuka1997} is
\begin{eqnarray}\label{EQN:muCRIT}
\mu\subC&=&\frac{2c\subS^2}{G}\;\,\simeq\;\,16.2\,{\rm M}_{_\odot}\,{\rm pc}^{-1}\,;
\end{eqnarray}
an isothermal filament with $\mu>\mu\subC$ should collapse and fragment. By this token, the L1495 Main Filament appears to be trans-critical, $\mu\sim\mu\subC$. (We note that an isothermal filament with $\mu<\mu\subC$ should relax towards hydrostatic equilibrium with a Plummer-like density profile and $p\!=\!4$ \citep{Ostriker1964} rather than $p\!=\!2$. We return to this issue in Section \ref{SEC:p4exp}.)

We can compare the variation of line-density along the L1495 Main Filament with the distribution of prestellar and protostellar cores. The black triangles and dashed vertical lines on Fig.~\ref{FIG:SegmentProps}(d) mark the locations of 28 {\it prestellar} cores from \citet{Marsh2015}; there are 19 on the B211 sub-filament, but only 9 on the B213 sub-filament. On Fig.~\ref{FIG:tauO}, the locations of the {\it protostellar} cores that had to be masked out before applying \ppp are shown with white circles; there are none associated with B211, but 9 on or near B213. 

To explain these variations, we presume that the L1495 Main Filament has accumulated -- and continues to accumulate -- mass from a turbulent inflow \citep{Clarke2017}. Its line-density has therefore never been uniform, and its rate of growth has varied with time. Some sections have become locally super-critical sooner than others, and some may thus far have always been sub-critical. Fragmentation has occurred where the filament has become locally super-critical \citep{Clarke2017,Chira2018}, spawning prestellar cores, which then condense into protostars, on a timescale $\,\sim0.5\,{\rm Myr}$ \citep[e.g.][]{Enoch2008}. Given the estimated flow-rate onto the L1495 Main Filament, $\sim 32\,{\rm M_{_\odot}\,pc^{-1}\,Myr^{-1}}$ \citep{Clarke2016,Palmeirim2013}, this condensation timescale is comparable with the timescale on which the critical line-density is replenished.

The inference is that, on average, B213 initially grew somewhat faster, became super-critical somewhat sooner, and fragmented into prestellar cores somewhat earlier, so that some of those prestellar cores have by now had time to become protostars; B213 is now in the process of being replenished, and will soon become supercritical again. In contrast, B211 has grown somewhat more slowly, and only became super-critical more recently; B211 contains prestellar cores, but none of them have yet evolved into protostars, so it is still marginally supercritical. 

In order to test this inference further, we have, for each of the 24 prestellar cores in or near the L1495 Main Filament, identified the nearest segment of the filament and noted the line-density of the associated {\it Segment Average Profile}. The red curve on Fig.~\ref{fig:coreKDE} shows the kernel-smoothed PDF of these line-densities, and the red vertical dashed line shows its median, $25.5\,{\rm M_{_\odot}\,pc^{-1}}$; in this averaging process, the contribution from a {\it Segment} is weighted by the number of prestellar cores to which it is the nearest {\it Segment}, so a few {\it Segments} are counted twice. For comparison, the blue curve and blue vertical dashed line show the kernel-smoothed PDF and median, $16.8\,{\rm M_{_\odot}\,pc^{-1}}$, of the line-densities of the remaining 72 {\it Segment Average Profiles}. By performing a Kolmogorov-Smirnov (KS) two-sample test, we show that the distance between the two distributions is $D_{_{\rm KS}} = 0.58$, and the probability that they are drawn from the same underlying distribution is $p_{_{\rm KS}} = 4 \times 10^{-6}$. Therefore the prestellar cores do indeed appear, almost exclusively, to lie on or near parts of the filament with supercritical line-density. 

\begin{figure*}
\centering
\includegraphics[width=1.0\linewidth]{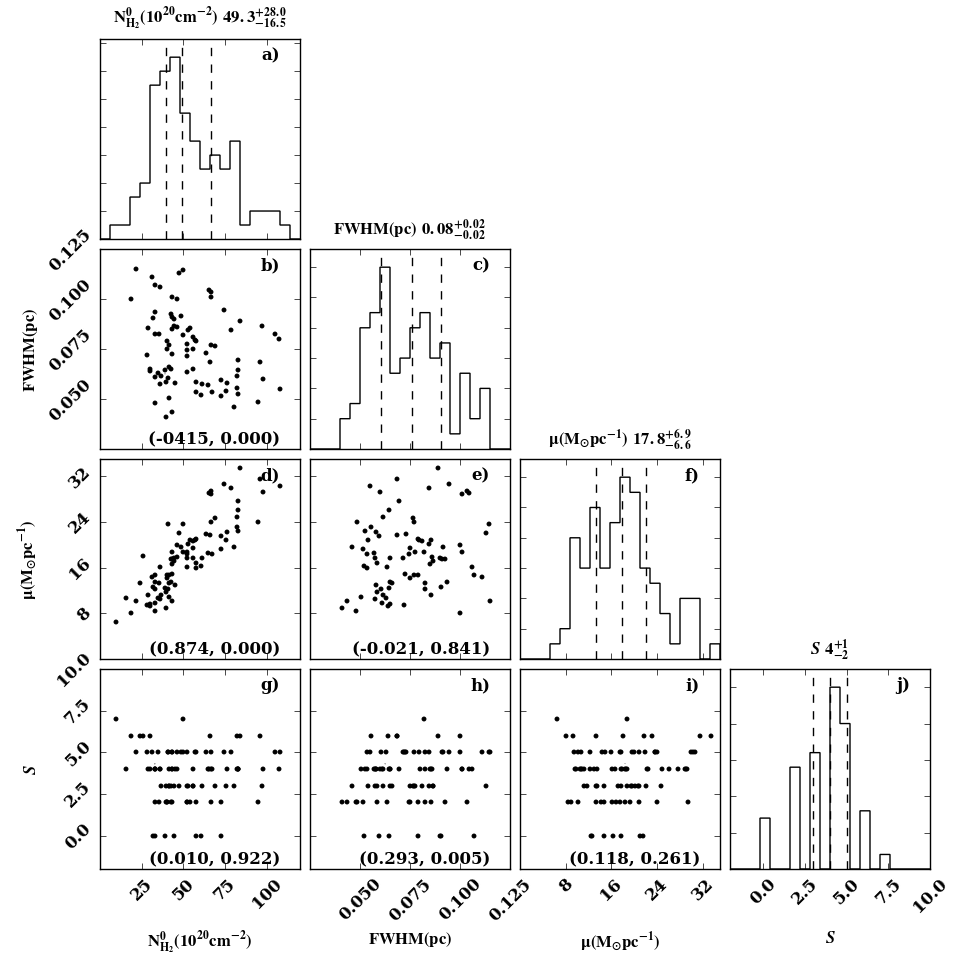}
\caption{Reading from top left to bottom right, the panels on the diagonal show the distributions of (a) $N\subO$, (c) {\sc fwhm}, (f) $\mu$ and (j) ${\cal S}$ for the segments along the L1495 Main Filament; the text above each of these panels gives the identity of the parameter, its mean value, and its interquartile range. The panels below, and to the left of, the diagonal give the correlations between the parameters, i.e. (b) $(N\subO,\mbox{\sc fwhm})$, (d) $(N\subO,\mu)$, (e) $(\mbox{\sc fwhm},\mu)$, (g) $(N\subO,{\cal S})$,  (h) $(\mbox{\sc fwhm},{\cal S})$, (i) $(\mu,{\cal S})$; the text at the bottom of each panel gives the Pearson correlation coefficients, in the format $(r_{_{\rm P}},p_{_{\rm P}})$.}
\label{FIG:Correls}
\end{figure*}

Fig. \ref{FIG:Correls} shows the distributions of, and correlations between, the parameters characterising the filament locally, i.e. $(N\subO,\mbox{\sc fwhm},\mu,{\cal S})$. $\mu$ and $N\subO$ are strongly correlated, while there is no apparent correlation between $\mu$ and $\mbox{\sc fwhm}$; therefore higher than average line-density is largely attributable to higher column-density rather than higher filament width. The parameter ${\cal S}$ is defined in Section \ref{SEC:sub}.

%%%%%
%%%%%

%%%%%
\subsection{Variation along the L1495 Main Filament, setting \texorpdfstring{$\boldsymbol{p=4}$}{p=4}}\label{SEC:p4exp}
%%%%%

We note that the contrast in column-density between the spine of the filament and the background is only $N\subO/N\subB\sim 20$. With this limited dynamic range, Plummer-like fits are affected by a degeneracy whereby small $p$ and small $r\subO$ are hard to distinguish from large $p$ and large $r\subO$\citep{Suri2018}. We have therefore re-fit the {\it Segment Average Profiles} setting $p\!=\!4$, which is the value expected for an isothermal cylinder in hydrostatic equilibrium \citep{Ostriker1964}. The resulting fits turn out to be better (reduced $\chi^2_{_{p=4}}=8.35$) than those we obtained with $p\!=\!2$ ($\chi^2_{_{p=2}}=29.23$). This in turn suggests that the L1495 Main Filament might be close to hydrostatic equilibrium --- in the sense that (a) the gas in the filament is approximately isothermal \citep[due to efficient CO cooling just inside the low-velocity accretion shock at the filament boundary,][]{WhitJaff2018}, and (b) the gas has had sufficient time to start to relax hydrodynamically (Whitworth, in prep.).

The $p\!=\!4$ fits give a scale-length, ${\bar r}\subO=0.063\pm 0.023\,{\rm pc}$ (as compared with ${\bar r}\subO=0.023\pm 0.006\,{\rm pc}$ for the $p\!=\!2$ fits), and hence $\overline{\textsc{fwhm}}=2(2^{2/3}-1)^{\!1/2}{\bar r}\subO=0.097\pm 0.036\,{\rm pc}$ (as compared with $\overline{\textsc{fwhm}}=12^{\!1/2}{\bar r}\subO=0.080\pm 0.020\,{\rm pc}$ for the $p\!=\!2$ fits). (The ratio $\textsc{fwhm}/r\subO$ is smaller for a $p\!=\!4$ fit than for a $p\!=\!2$ fit, because the $p\!=\!4$ Plummer-like profile drops much more abruptly outside $r\!=\!r\subO$ than the $p\!=\!2$ profile.)

There might appear to be a contradiction here. When we fit the {\it Segment Average Profiles} with $p\!=\!4$, we get a better fit than with $p\!=\!2$. However, when we fit the {\it Global Average Profile} with $p$ treated as a free parameter we get the best fit with $p\!=\!2.02$ on the North side of the filament, and $p\!=\!1.73$ on the South side, i.e. much closer to $p\!=\!2$. The reasons are twofold. First, in delineating the spine (Section \ref{SEC:cyl}) it gets displaced from the true column-density maxima by the smoothing. Second, and more importantly, the {\it Global Average Profile} is basically a sum of 1032 individual {\it Local Sample Profiles}, each with its own central column-density, $N\subO$, and scale-length, $r\subO$; this has the effect of broadening and flattening the {\it Global Average Profile}, and hence reducing the apparent $p$ value.

%%%%%
\section{Internal substructure within the L1495 Main Filament}\label{SEC:sub}
%%%%%

The Plummer-like fits obtained in the previous section (Section \ref{SEC:cyl}) should be viewed as azimuthally averaged profiles, in which any internal sub-structure has been smoothed out. There is evidence from observations of the C$^{18}$O ($J\!=\!1\!-\!0$) line \citep{Hacar2013} that the L1495 Main Filament has significant internal substructure. Specifically \citet{Hacar2013} find evidence for coherent, plaited `fibres' in their position-position-velocity (PPV) data-cubes. The interpretation of these features is controversial, with \citet{Clarke2018} warning that `fibres' are not necessarily linked to coherent 3-D structures. Detailed simulations of the assembly of a filament from a turbulent inflow \citep{Clarke2017} shows (a) that the filament tends to break up into sub-filaments, and (b) that the velocity dispersion between these sub-filaments (a very non-isotropic macro-turbulence) can delay the overall collapse of a filament whose line-density already exceeds the critical value, $\mu\subC$ (Eqn. \ref{EQN:muCRIT}), for a filament in hydrostatic equilibrium. 

The {\sc FilChaP} algorithm \citep{Suri2018} returns a parameter ${\cal S}$, which is a measure of substructure within the filament. Specifically, ${\cal S}$ is the number of secondary peaks (maxima) and shoulders (points of inflexion), resolved by at least five pixels, that remain once a Plummer-like profile has been fitted to the primary peak of the {\it Segment Average Profile}. Since here ${\cal S}$ is obtained from a 2D map of column-density, we presume that some 3D substructure is lost due to projection, and therefore we interpret ${\cal S}$ as a lower limit on the amount of true 3D internal substructure.

Fig. \ref{FIG:Correls} shows that there is no statistically significant correlation between ${\cal S}$ and the column-density through the spine, $N\subO$, with Pearson correlation parameters \citep{Pearson1895} $(r\subP,p\subP)=(0.010,0.922)$; here $r\subP$ is the Pearson correlation coefficient, and $p\subP$ is the p-value, or probability of finding $r\subP$ given a data set with no correlation. There is a weak correlation between ${\cal S}$ and the full width at half maximum, {\sc fwhm}, with $(r\subP,p\subP)=(0.293,0.005)$, i.e. fatter sections of the filament seem to have more resolved substructure. There is no statistically significant correlation between ${\cal S}$ and the line-density, $\mu$, with $(r\subP,p\subP)=(0.118,0.261)$.

%%%%%
\section{Discussion}\label{SEC:disc}
%%%%%

%%%%%
\subsection{Comparison with Herschel observations}\label{SEC:synthObs}
%%%%%

\begin{figure*}
\centering
\includegraphics[width=\textwidth]{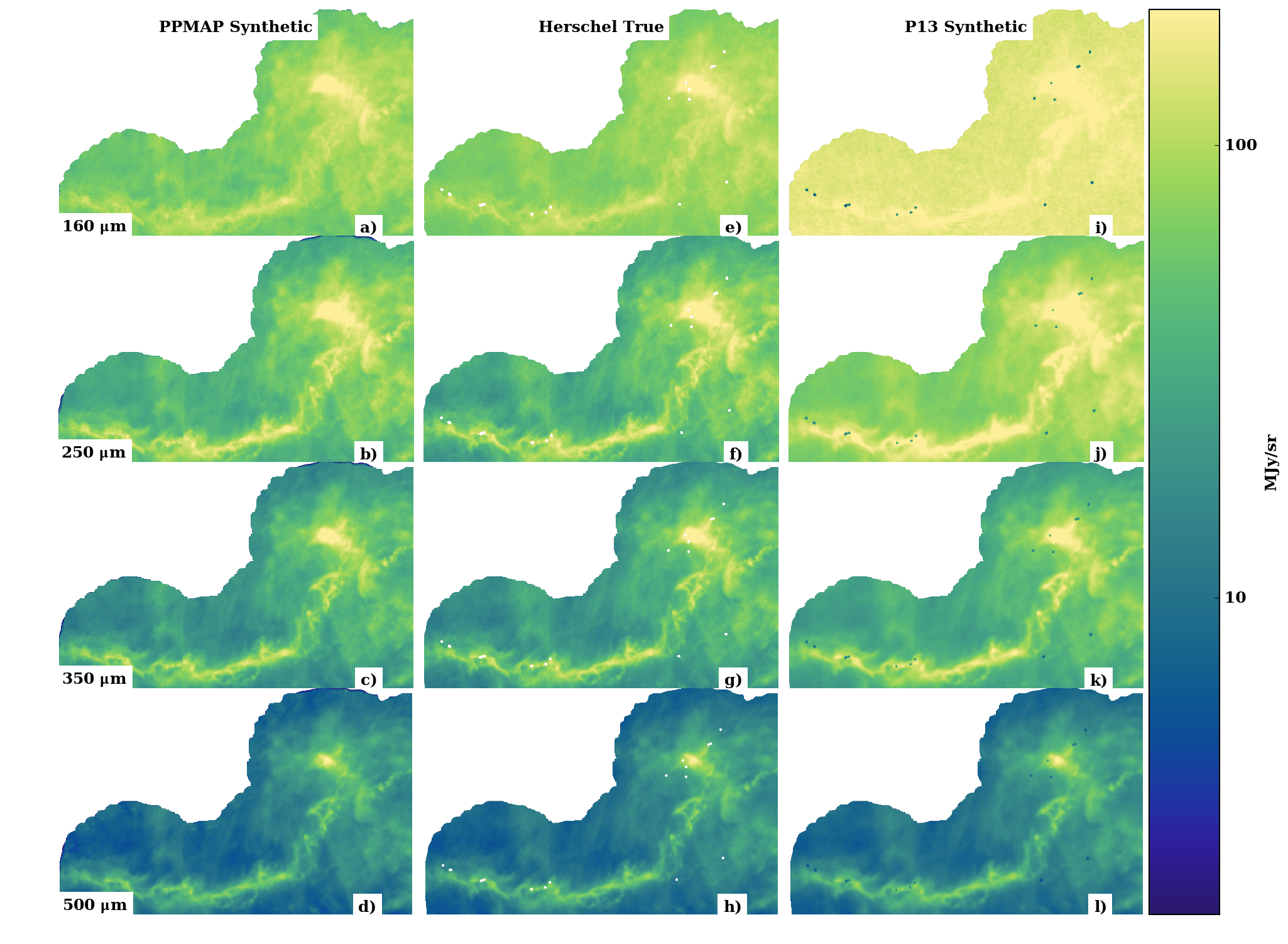}
\caption{Panels in the lefthand and righthand columns show synthetic maps in four different {\it Herschel} bands, derived from, respectively, the \ppp and the \citetalias{Palmeirim2013} results; panels in the central column show the corresponding true {\it Herschel} maps. From top to bottom the different rows correspond to Herschel wavebands at \SIlist{160;250;350;500}{\micro \meter}.}
\label{FIG:synthHerc}
\end{figure*}

To assess the fidelity of the \ppp results, and to compare them with those obtained by \citetalias{Palmeirim2013}, we have computed synthetic maps, and compared them with the original {\it Herschel} maps (see Fig. \ref{FIG:synthHerc}). For the \ppp data products the monochromatic intensity, $I_{_\lambda}$, is given by Eqn. (\ref{EQN:PPInt}), and for the \citetalias{Palmeirim2013} data products by Eqn. (\ref{EQN:standInt}). These intensities are then convolved with the spectral response functions of the different {\it Herschel} wavebands, and with their beam profiles. Finally they are re-gridded to match the pixel sizes of the original {\it Herschel} observations in each band, and colour correction factors are applied so as to negate the corrections applied during the column-density fitting process.

The lefthand column of Fig. \ref{FIG:synthHerc} shows the synthetic maps computed in this way from the \ppp data products. The righthand column shows the synthetic maps computed in the same way from the \citetalias{Palmeirim2013} data products. And the central column shows the original {\it Herschel} maps. All maps are masked to exclude areas outside the SCUBA-2 pointings. Reading from the top, the maps are for the {\it Herschel} \SIadj{160}{\micro \meter}, \SIadj{250}{\micro \meter}, \SIadj{350}{\micro \meter} and \SIadj{500}{\micro \meter} wavebands; maps for the \SIadj{70}{\micro \meter} and \SIadj{850}{\micro \meter} wavebands have not been presented since \citetalias{Palmeirim2013} do not utilise these wavebands. Across all four {\it Herschel} wavebands the \pp-based synthetic maps match the original {\it Herschel} maps better than the \citetalias{Palmeirim2013}-based maps.

%%%%%
%%%%%

To quantify the quality of the synthetic maps in the different wavebands, we compute a goodness of fit parameter, 
\begin{eqnarray}\label{EQN:G}
\mathcal{G}_{_{\cal B}} = \left(\frac{\sum\limits_{_{\rm PIXELS}}\!\left\{I_{_{\rm{\cal B}.true}}^{\,-1}\,\left(I_{_{\rm{\cal B}.synth}}-I_{_{\rm{\cal B}.true}}\right)^2   \right\}}{\sum\limits_{_{\rm PIXELS}}\!\left\{I_{_{\rm{\cal B}.true}}\right\}}\right)^{1/2}\label{EQN:SynthHercComp}
\end{eqnarray}
Here, $I_{_{\rm{\cal B}.true}}$ is the observed true intensity in band ${\cal B}$, and $I_{_{\rm{\cal B}.synth}}$ is the computed synthetic intensity in band ${\cal B}$. Thus ${\cal G}_{_{\cal B}}$ is the root mean square fractional difference between the synthetic and true maps, averaged over all the pixels, with the contribution to the mean from each pixel weighted by its true intensity. A good match between the two maps will yield a small ${\cal G}_{_{\cal B}}.\;$ Table \ref{tab:goodness} gives values of ${\cal G}_{_{\cal B}}$ for the four {\it Herschel} wavebands that are common to this study and to \citetalias{Palmeirim2013}. Values are given both for the entire map, and for a more restricted area defined by excluding all pixels that are more than $0.5\,{\rm pc}$ from the spine of the L1495 Main Filament (this removes edge effects and the hotter region close to V892 Tau). In all cases, \ppp reproduces the {\em Herschel~} observations much more accurately than \citetalias{Palmeirim2013}, particularly at shorter wavelengths.

%%%%%
\begin{table}
\centering
\begin{tabular}{ccccc}\hline
 & \multicolumn{2}{c}{{\sc Whole Map}} & \multicolumn{2}{c}{{\sc Filament Only}} \\
{\sc Band} & ${\cal G}_{_{{\cal B}.\rm PPMAP}}$ & $ {\cal G}_{_{{\cal B}.\rm P13}}\;\;$ & ${\cal G}_{_{{\cal B}.\rm PPMAP}}$&${\cal G}_{_{{\cal B}.\rm P13}}\;\;$ \\
\SI{160}{\micro \meter} & 0.23 & 1.05 & 0.13 & 1.01 \\
\SI{250}{\micro \meter} & 0.11 & 1.02 & 0.10 & 0.92 \\
\SI{350}{\micro \meter} & 0.08 & 0.37 & 0.07 & 0.38 \\
\SI{500}{\micro \meter} & 0.08 & 0.09 & 0.08 & 0.10 \\\hline
\end{tabular}
\caption{Goodness of fit parameter, $\mathcal{G}_{_{\cal B}}$ (Eqn. \ref{EQN:G}), for synthetic observations derived from the \ppp and P13 column-density and temperature estimates. Values are given both for the whole map (columns 2 \& 3), and for a $1\,{\rm pc}$-wide strip surrounding the L1495 Main Filament (columns 4 \& 5).}
\label{tab:goodness}
\end{table}
%%%%%

%%%%%
\subsection{Mantle growth}\label{subsec:growth}
%%%%%

Since we have found evidence that the dust in the filament is different from that in the suroundings, we should consider whether this is feasible. To estimate how long it takes for grains to accrete mantles, we consider a generic neutral gas-phase molecule with mass $m_{_{\rm MOL}}\!=\!\mu_{_{\rm MOL}}m_{_{\rm H}}$ and arithmetic mean speed ${\bar v}_{_{\rm MOL}}\!=\!\left(8k_{_{\rm B}}T/\pi m_{_{\rm MOL}}\right)^{1/2}$; here $T$ is the gas-kinetic temperature, as distinct from the dust temperature, $T_{_{\rm D}}$. If a representative spherical dust grain has radius $r_{_{\rm D}}\!=\!10^{-5}\,{\rm cm}$ and internal density $\rho_{_{\rm D}}\!=\!3\,{\rm g}\,{\rm cm}^{-3}$, then (using the parameters defined in Section \ref{SEC:ConvFact}), the number-density of dust grains is $n_{_{\rm D}}\!=\!6Z_{_{\rm D}}n_{_{\rm H_2}}m_{_{\rm H}}/4\pi r_{_{\rm D}}^3\rho_{_{\rm D}}X$ and the geometric cross-section of a dust grain is $\varSigma_{_{\rm D}}\!=\!\pi r_{_{\rm D}}^2$. Hence the rate at which the generic neutral gas-phase molecule strikes a dust grain is ${\cal R}_{_{\rm HIT}}\!=\!n_{_{\rm D}}\varSigma_{_{\rm D}}{\bar v}_{_{\rm MOL}}$. The accretion timescale for the generic neutral gas-phase molecule is therefore
\begin{eqnarray}\nonumber
t_{_{\rm ACC}}\!\!&\!\!=\!\!&\!\!\left({\cal R}_{_{\rm HIT}}{S}\right)^{-1}\;=\;\frac{4Xr_{_{\rm D}}\rho_{_{\rm D}}}{6m_{_{\rm H}}Z_{_{\rm D}}n_{_{\rm H_2}}}\left(\frac{\pi m_{_{\rm MOL}}}{8k_{_{\rm B}}T}\right)^{1/2}\frac{1}{S}\\\label{EQN:tACC}
\!\!&\!\!\simeq\!\!&\!\!0.3\,{\rm Myr}\left(\!\frac{n_{_{\rm H_2}}}{\rm 10^4H_{_{2}}cm^{-3}}\!\right)^{\!-1}\!\!\left(\!\frac{T}{10\,{\rm K}}\!\right)^{\!-1/2}\!\!\left(\!\frac{\mu_{_{\rm MOL}}}{28}\!\right)^{\!1/2}\frac{1}{S}.\hspace{0.8cm}
\end{eqnarray}
Here $\mu_{_{\rm MOL}}\!=28$ corresponds to the $^{12}{\rm C}^{16}{\rm O}$ molecule, and ${S}$ is the sticking coefficient, normally assumed to be of order unity at low temperatures \citep[$T\la$\SI{40}{\kelvin};][]{He2016}. For comparison, the timescale for spherical freefall collapse is 
\begin{eqnarray}\label{EQN:tFF}
t_{_{\rm FF}}\!\!&\!\!=\!\!&\!\!\left(\frac{3\pi X}{64Gm_{_{\rm H}}n_{_{\rm H_2}}}\right)^{1/2}\,\simeq\,0.3\,{\rm Myr}\,\left(\!\frac{n_{_{\rm H_2}}}{\rm 10^4H_{_{2}}cm^{-3}}\!\right)^{\!-1/2}\!.\hspace{0.8cm}
\end{eqnarray}
Since freefall collapse requires that resistance due to thermal, turbulent and magnetic pressure is negligible, $t_{_{\rm FF}}$ is probably the minimum timescale on which the density can increase; a shorter timescale would require an implausibly large and well focussed inward ram pressure. Therefore, at the high densities, $n_{_{\rm H_2}}\,\ga\,10^4\,{\rm H_{_2}\,cm^{-3}}$, obtaining near the centre of the L1495 Filament, mantle growth is likely.

%%%%%
\section{Conclusions}\label{SEC:conc}
%%%%%

We have re-analysed {\em Herschel} observations of the L1495 Main Filament, using the new \ppp procedure. \ppp returns a 4D data-cube giving, for each pixel on the sky, the column-density of dust of different types (different emissivity indices, $\beta$) and at different temperatures ($T\subD$). The \ppp results indicate that previous estimates of the width of the filament ({\sc fwhm}), and of its line-density ($\mu$) need to be revised to significantly lower values. They also provide evidence that the interior of the filament is cooler than the outer layers, and that the dust in the interior has different properties from that in the surroundings. Specifically \ppp reveals the following features of the L1495 Main Filament.

\begin{enumerate}%[i]
\item Most of the dust within the L1495 Main Filament has temperature in the range \SIrange{8}{16}{\kelvin}.
\item Much of the diffuse dust outside the L1495 Main Filament has temperature in the range \SIrange{16}{18}{\kelvin}.
\item \ppp returns a broader range of dust temperatures than the standard procedure.
\item Most of the dust in the interior of the L1495 Main Filament has emissivity index $\beta\la 1.5$.
\item Most of the diffuse dust outside the L1495 Main Filament has emissivity index $\beta\ga 1.7$.
\item We have shown (Section \ref{subsec:growth}) that the volume-density in the interior of the L1495 Filament is sufficiently high that significant grain growth by mantle accretion is possible, i.e. the timescale for mantle accretion is probably shorter than the dynamical timescale.
\item The {\it Global Average Profile} of the whole L1495 Main Filament is well fit by a Plummer-like density distribution (Eqn. \ref{EQN:rhor}), with central density $n\subO\sim 2.8\times 10^4\,{\rm H_{_2}\,cm^{-3}},\;$ scale-length $r\subO\sim 0.022\,{\rm pc},\;$ and $p\sim 1.88.\;$ With this density profile, the column-density has $\textsc{fwhm}\sim 0.087\,{\rm pc}$.
\item Although this $\textsc{fwhm}$ appears to very close to the value $0.09\,{\rm pc}$ reported by \citet{Arzoumanian2011} and \citetalias{Palmeirim2013}, we note that if we had evaluated $\textsc{fwhm}$ in the same way they did we would have obtained a rather smaller value, $0.056\,{\rm pc}$.
\item The rather shallow radial density gradient of the {\it Global Average Profile} (i.e. Plummer-like exponent $p\!\sim\!2$) is a consequence of the smoothing and averaging inherent in its derivation. When we analyse small local segments of the filament (of length $0.004\,{\rm pc}$), they are better fit with $p\!=\!4$, implying that -- if we assume the gas is isothermal -- they may be close to hydrostatic equilibrium \citep{Ostriker1964}.
\item The line-density of the filament, $\mu$, has a median and interquartile range of $\;17.6^{+6.9}_{-6.7}\,{\rm M_{_\odot}\,pc^{-1}}$.
\item If we adopt the canonical value for the critical line-density above which an isothermal molecular filament at $10\,{\rm K}$ cannot be supported against self-gravity by a thermal pressure gradient, i.e. $\mu\subC\!=\!16\,{\rm M_{_\odot}\,pc^{-1}}$, only local sections of the L1495 Main Filament are presently unstable against collapse and fragmentation.
\item Sections of the L1495 filament that are super-critical ($\mu>\mu\subC$) tend to coincide with the locations of prestellar cores. This is compatible with the plausible scenario in which it is the local line-density, rather than the global line-density, that determines whether, and where, a filament is unstable against fragmentation.
\end{enumerate}

%%%%%
\section*{Acknowledgements}
%%%%%

ADPH gratefully acknowledges the support of a postgraduate scholarship from the School of Physics \& Astronomy at Cardiff University and the UK Science and Technology Facilities Council. APW, MJG and ODL gratefully acknowledge the support of a consolidated grant (ST/K00926/1), from the UK Science and Technology Funding Council. ODL is also grateful for the support of an ESA fellowship. SDC gratefully acknowledges support from the ERC starting grant No. 679852 `RADFEEDBACK'. We thank S{\"u}meyye Suri for her help in extracting continuous structures from the DisPerSE algorithm, Emily Drabek-Maunder for providing the SCUBA-2 observations, and Philippe Andr\'e for providing very helpful feedback on an earlier version. We also thank the referee for their encouraging report. The computations have been performed on the Cardiff University Advanced Research Computing facility, \verb|ARCCA|.

%%%REFERENCES %%
\bibliographystyle{mnras}
\bibliography{L1495PPMAP}
% Don't change these lines
\bsp	% typesetting comment
\label{lastpage}
\end{document}